\documentclass[nonblindrev]{informs3} 

\SingleSpacedXI 

\usepackage{bm}
\usepackage[ruled,commentsnumbered]{algorithm2e}
\usepackage{algpseudocode}
\usepackage{graphicx}
\usepackage{caption}
\usepackage{subcaption}

\usepackage{endnotes}
\let\footnote=\endnote

\usepackage{natbib}
 \bibpunct[, ]{(}{)}{,}{a}{}{,}%
 \def\bibfont{\small}%

\TheoremsNumberedThrough     
\ECRepeatTheorems

\EquationsNumberedThrough    

\begin{document}

\RUNAUTHOR{Georgiadis, Iosifidis and Tassiulas}

\RUNTITLE{Sharing Economy Networks}

\TITLE{On the Efficiency of Sharing Economy Networks}

\ARTICLEAUTHORS{%
\AUTHOR{Leonidas Georgiadis}
\AFF{Department of Electrical and Computer Engineering, Aristotle University of Thessaloniki, Greece, \EMAIL{leonid@auth.gr}} 
\AUTHOR{George Iosifidis}
\AFF{Department of Computer Science and Statistics, Trinity College Dublin, Ireland, \EMAIL{george.iosifidis@tcd.ie}}
\AUTHOR{Leandros Tassiulas}
\AFF{Department of Electrical Engineering and Institute for Network Science, Yale University, USA, \EMAIL{leandros.tassiulas@yale.edu}}
}

\ABSTRACT{We consider a sharing economy network where agents embedded in a graph share their resources. This is a fundamental model that abstracts numerous emerging applications of collaborative consumption systems.	The agents generate a random amount of spare resource that they can exchange with their one-hop neighbors, seeking to maximize the amount of desirable resource items they receive in the long run. We study this system from three different perspectives: a) the central designer who seeks the resource allocation that achieves the most fair endowments after the exchange; b) the game theoretic where the nodes seek to form sharing coalitions within teams, attempting to maximize the benefit of their team only; c) the market where the nodes are engaged in trade with their neighbors trying to improve their own benefit. It is shown that there is a unique family of sharing allocations that are at the same time most fair, stable with respect to continuous coalition formation among the nodes and achieving equilibrium in the market perspective. A dynamic sharing policy is given then where each node observes the sharing rates of its neighbors and allocates its resource accordingly. That policy is shown to achieve long term sharing ratios that are within the family of equilibrium allocations of the static problem. The equilibrium allocations have interesting properties that highlight the dependence of the sharing ratios of each node to the structure of the topology graph and the effect of the isolation of a node on the benefit may extract from his neighbors.}%

\maketitle

%


\section{Introduction}

\subsection{Motivation}

Collaborative consumption \citep{felson-coco-book}, or the \emph{sharing economy}, is an emerging economic trend that promotes novel models of sharing, bartering, or renting resources and services, which is opposed to traditional ownership-based models \citep{botsman-coco-book,sundarajan}. These solutions are attracting increasing interest due to the global recession that has changed the consumer behavior, the pressing environmental concerns, and the penetration of Internet that facilitates such activities \citep{nytimes-sharing,economist-13,guardian-14}. The success of sharing economy is best manifested by the fact that it already encompasses a very diverse set of models. In some cases the payments are implemented with legal tender currency while in others the sharing activities are supported by bespoken credit systems; some applications have geographically-restricted scope while others operate in a world-wide scale; in many models the users' collaboration is mediated by third parties, as in the transportation network Uber, while in other cases the collaboration takes place directly among users who are co-located or have common interests. All these consumption-as-a-service schemes offer sustainable and low-cost solutions to daunting consumption problems, and can boost the economy at a local or larger scale \citep{PwC}.

The fundamental goal in sharing economy is to leverage the potential of ubiquitous connectivity and enable the exchange of resources among the users by exploiting the complementarity of their resource availability and demands. In decentralized implementations, this can be achieved as follows: whenever a user has some idle resource, she offers it to other users who at that time have excess needs, and benefits from the resources they offer to her in the future. Such solutions can also address problems where users have different preferences for the different resources, and need to exchange them in order each one to acquire the most valuable for her needs \citep{unver-book}. The common denominator in these scenarios is that users are both resource consumers and producers (or, prosumers), and they are free to decide how their resources will be allocated to others. Moreover, their collaboration opportunities are constrained by various network graphs. For example, ride sharing or food sharing is constrained by the geographic proximity of the participants, renewable energy sharing relies on the grid network, commodity exchanges are conditioned on the matching of the users\textquoteright{} needs or social ties, and so on. We will use hereafter the term \textit{sharing economy networks} to describe these models where users (or, agents) embedded in a network graph exchange their resources over time, aiming to maximize their individual benefits.

Despite the proliferation and huge potential of these sharing economy networks, very important questions about their salient features and performance remain unanswered. For example, to date it is not understood if these sharing systems admit \emph{sharing equilibriums} nor if there are meaningful dynamic sharing policies that can lead to these equilibriums. Also, it is not known how efficient these equilibriums are in terms of social welfare, and how robust to strategic behaviors of users who act independently or within coordinated groups. Besides, we cannot assess today how the network affects the overall performance of the sharing system or to what extent the network position of a user shapes the resources she will receive from the network. 

Motivated by these important questions, we consider a general model for sharing economy networks and follow a systematic approach to articulate and analyze the following issues:
\begin{itemize}
\item \underline{Definition and Properties of a Fair Sharing Policy.} This is one of the most critical issues in sharing economy networks. Ideally, from a system design point of view, each user should receive resource commensurate to its contribution. This is necessary to establish the sense of fairness in the participants. However, this is not always possible because of the underlying graph that prescribes, for each user, the subset of users she can collaborate with. Additionally, there may be multiple feasible sharing solutions that differ on the amount of resource each user receives. We would prefer to select among them a \emph{fair} outcome that balances the exchanges as much as possible. In the context of sharing economy networks, such a fair allocation is also considered efficient as it minimizes the wasted resources and maximizes the social welfare. The existence and the characterization of the properties of centrally designed fair and efficient policies (e.g., their dependency on the underlying network graph) is an important and currently open question.
	
\item \underline{Existence and Fairness of Sharing Equilibriums}. Most often these systems are not controlled by a central entity that can exogenously impose such a fair solution. Instead, each user is free to decide her strategy and therefore allocate her idle resource to the neighbors from which she receives more service in return. Such interactions give rise to barter-style competitive sharing markets where the users exchange their resources in a greedy fashion. The main question here is if these myopic strategies lead to an equilibrium allocation where each user cannot unilaterally improve her benefits, and whether these equilibriums are affected by the network constraints. In more advanced settings, groups of users might be able to form coalitions and exclude non-members from sharing. For example, in a Wi-Fi sharing community such as FON \citep{FON} a subset of users may decide to serve only each other, expecting to increase their own benefits. Such strategies are very likely to deteriorate the system's performance, e.g., resulting in isolated users, and it is important to explore if there are coalitional equilibriums that partition the network. Finally, a naturally arising question is how efficient these competitive or coalition equilibriums are, i.e., whether they are related to the above centrally designed fair sharing policy.
	
\item \underline{Dynamics of Sharing Interactions}. In such decentralized systems, the issue of how the users can reach the sharing equilibriums is equally important to the existence of the latter. Therefore, it is crucial to understand whether there are meaningful and simple-to-implement dynamic allocation rules that can enable users to exchange their randomly created resources in a fashion that is fair and efficient in the long-term. For the sharing market setting, such dynamic policies must also be incentive-compatible, i.e., aligned with the users' efforts to maximize their own benefits accrued from the sharing network. Besides, another desirable feature of these sharing policies is to rely on the minimum possible information regarding the structure of the network and the resources or decisions of the users. All these requirements sum up to the following key question: when the users devise their dynamic (over time) sharing decisions in a self-interested and myopic fashion, having information about the respective actions only of their one-hop neighbors, can the sharing network reach an equilibrium that is also fair and robust to group strategic deviations?

\end{itemize}

The problem of efficient cooperation and resource sharing, even when network constraints are involved, has been studied in several contexts. In communication and computing systems for example, many architectures rely on pooling of resources that belong to different businesses (e.g., Internet Service Providers) or even to different end-users. Examples of the latter cases include bandwidth management in peer-to-peer file sharing systems \citep{RJohariToNBilateral2011, zhang-proportional}, Wi-Fi Internet sharing communities \citep{Sofia-UPN,efstathiou}, online content distribution schemes \citep{ioannidis-peer-assisted}, and so on. Similar ideas have been explored in other contexts, e.g., for facilitating renewable energy sharing in smart grid networks \citep{saad-smart}, or for enabling operating costs' reduction in networks of service providers or manufacturers \citep{OR - Cooperation in Service,trading networks - manea,cost sharing - tobias}. However, prior works in those areas do not address the questions outlined above, namely the impact of the graph constraints and the analysis of the competitive interactions of the users over time. This latter element gives rise to a new type of bartering markets which are related to the general equilibrium theory and the seminal work of Arrow and Debreu \citep{arrow-debreu}. Here we study a different setting which, not only has graph constraints similarly to other networked economies \citep{KearnsEconSocial2004,KearnsGraphEcon2004}, but also does not presume the existence of a monetary instrument. Besides our focal point is the dynamics of agents' decisions, an aspect that remains under-investigated even in general competitive market models who rely on full-information tatonnement processes \citep{ColellWhinstonGreenBook1995}.

Finally, the emergence of sharing economy has motivated early studies focusing on specific applications such as vehicle pooling \citep{PNAS-vehicle-pooling} or sharing of mobile data plans \citep{laoutaris-conext}. Nevertheless, these studies do not answer the above fundamental questions arising in sharing economy. In the sequel we provide a comprehensive overview of the literature highlighting the recent studies about sharing economy, as well as related resource sharing models that have been proposed in pertinent areas.

\subsection{Methodology and Contributions}

In order to shed light on these questions, we employ a general model that captures many basic instances of the emerging sharing economy networks. We consider a set of users, where each one generates over time a random amount of idle resource that she does not need and therefore can allocate to her neighbors. For instance a user may have excess bandwidth in her connection plan, that may either being used to forward some neighbor's traffic or get wasted. We assume that each user has unsaturated demand for the resources of others. This model captures situations where the users have complementary resource availability and demand generation, or different types of resources and different preferences for them. Neighborhood relationships are described by a bidirectional or directional connected graph which does not change. The \emph{sharing ratio} (or, simply \emph{ratio}) of total received over allocated long term average resource characterizes the performance of each user, as it quantifies the accrued benefits over her contributions. We consider that the resources can be shared only among one-hop neighbors, and the shared resources are directly
consumed by their recipients and cannot be distributed further in the network. This assumption captures practical distribution constraints that arise in many sharing systems.

From a system point of view, a central designer would prefer to have a vector of sharing ratios where each coordinate that corresponds to a user (or, a node in the graph), has value equal to one. Often this will not be possible due to the graph exchange constraints and asymmetries in nodes' resource availability. For example, in a microgrid energy sharing network, some renewables might create very large amounts of energy which cannot be matched by the neighboring devices. For these cases, the lexicographically maximum (lex-optimal), or max-min, sharing vector is a meaningful performance criterion as it is Pareto optimal and balances the shared resources as much as possible \citep{nace-tutorial}.

In the absence of a network controller however, we assume that each node makes greedy myopic allocation decisions so as to maximize the aggregate resource it receives in return from the community. The interactions of the nodes give rise to a competitive market, which however differ from previous similar models  \citep{arrow-debreu, zhang-proportional} due to the existence of the graph and the absence of side-payments (money) among the nodes. We introduce the concept of sharing equilibrium that is appropriate for this setting, characterize the equilibrium allocations, and study its relation to the centrally designed max-min fair policy.

Accordingly, we assume that subset of nodes can coordinate and form coalitions exchanging resources only with each other. A coalitional graph-constrained game with non-transferable utility (NTU) is identified in the above set-up. We focus on the existence and properties of stable equilibrium allocations. Given a certain global allocation, if there is a subset of nodes that when they reallocate their own resources among themselves manage to improve the sharing ratio of \emph{at least one} node in the subset, then they have an incentive to deviate from the global allocation. Therefore, when an allocation is in equilibrium, it should be \emph{strongly stable} and no such subset should exist.

We study the above frameworks, that differ on the assumptions about the system control and the users behavior, and find a surprising connection among them. In particular:

(\textbf{i}) It is proved that there is a unique sharing equilibrium ratio vector that is a solution for the competitive market, and lies in the core of the NTU graph-constrained coalitional game, being also strongly stable. This is the max-min fair ratio vector. This result reveals that a centrally designed meaningful fair solution is robust to nodes' selfish strategies even if they are allowed to coordinate and form strategic groups seeking to improve their payoff. This finding has many implications for the applicability of such fair policies to sharing economy systems.

(\textbf{ii}) It is shown that the equilibrium exhibits rich structure and a number of interesting properties. For example, in the equilibrium allocation there is exchange of resources only among the nodes with the lowest sharing ratios and the nodes with the highest ratios, the nodes with the second lowest ratios with the set of the second highest ratios, and so on. We also study how the sharing ratios are affected by the graph properties, such as the node degree. This latter aspect is particularly important from a network design point of view as it reveals, among others, the impact a link removal or addition has on the equilibrium. Our findings hold for any graph, and therefore they can help a controller to predict or even dictate the sharing equilibrium.

(\textbf{iii}) Finally, we propose a distributed stochastic algorithm that can be used by the nodes in order to make sharing decisions over time. The algorithms is simple and with minimal information requirements as it allocates the resource generated at each time instance at a node to its neighbor having the highest exchange ratio at that point. This strategy is intuitive as well, since it maximizes the current sharing benefits for the users. Interestingly, it is proved that this dynamic algorithm leads to the above fair and robust sharing equilibrium points.

The rest of this paper is organized as follows. In Section \ref{sec:Model-Notation} we present the model and the problem statement ; in Section \ref{sec:algorithms} we introduce a policy that solves the problem for all three frameworks; Section \ref{sec:Numerical-Results} presents extensive numerical results and Section \ref{sec:Related} surveys the related literature in different areas. We conclude in Section \ref{sec:conclusions} where we also discuss our model assumptions. All the proofs can be found in the Appendix of the paper.

\section{Model and Problem Statement} \label{sec:Model-Notation}

\subsection{Notation and Model}

We use capital letters to denote sequences of random variables, e.g., $\left\{ X_{i}\right\} ,$ or $\left\{ X\left(t\right)\right\}$. Time averages of sequences are denoted with the same letter and a bar on top, e.g., 
\[
\bar{X}\left(t\right)=\frac{\sum_{\tau=1}^{t}X\left(\tau\right)}{t}.
\]
Let $G=(\mathcal{{N}},\mathcal{{E}})$ denote\textcolor{black}{{} a connected undirected} graph with a set $\mathcal{N}$ of $N=|\mathcal{\mathcal{N}}|$ nodes and a set $\mathcal{E}\subseteq\left\{ \left(i,j\right):\ i,j\in\mathcal{N},\ i\neq j\right\} $ of $E=|\mathcal{{E}}|$ links. We denote by $\mathcal{N}_{i}$ the set of neighbors of node $i$, that is $\mathcal{N}_{i}=\left\{ j:\ (i,j)\in\mathcal{E}\right\}$. We consider a system that evolves over time and we assume a slotted time operation, where slot $t=1,2,\cdots,$ is the time interval $[t,t+1).$ The ``beginning'' and ``end'' of slot $t$ are respectively the times $t$ and $t+1$. The dynamics of the nodes interactions can be described as follows. At the beginning of time slot $t$, node $i$ generates resource $D_{i}(t)$, where $D_{i}\left(t\right),\ t=1,2,...,$ are i.i.d with mean $E\left[D_{i}\left(t\right)\right]=D_{i}>0$; to avoid complications in the discussion we will assume that $D_{i}\left(t\right)$ are bounded, i.e., there is a real number $B$ such that $D_{i}\left(t\right)\leq B,\ i\in\mathcal{N},\ t=1,2,\cdots.$ The long-term average amount of produced resource by node $i$, $D_{i}$,
will be referred to as ``endowment'' of node $i$. This resource is distributed to the neighbors of $i$ according to a policy $\pi$
which is formally defined below. 
\begin{definition} 
A policy $\pi$ is a set of rules according to which the distribution of resources among the nodes in $\mathcal{N}$ is effected over time. More specifically, a policy $\pi$ determines the amount or resource node $i$ gives to node $j\in\mathcal{N}_{i}$ at time $t,$ based on the generated and allocated resources up to time $t.$ We denote the class of all policies by $\Pi.$
\end{definition} 
Under a policy $\pi$, at time $t$ node $i$ gives to node $j$ amount $D_{ij}^{\pi}\left(t\right)\geq0$ of the resource it generates up
to $t,$ and since the node cannot give more than it generates, it holds for any $t$, 
\begin{equation}
\sum_{\tau=1}^{t}\sum_{j\in\mathcal{N}_{i}}D_{ij}^{\pi}(\tau)\leq\sum_{\tau=1}^{t}D_{i}(\tau).\label{eq:resource-allocated}
\end{equation}
The average amount of resource node $i$ gives to its neighbors by time $t$is 
\[
\bar{D}_{i}\left(t\right)=\frac{\sum_{\tau=1}^{t}\sum_{j\in\mathcal{N}_{i}}D_{ij}^{\pi}(\tau)}{t}
\]
The amount of resource node $i$ receives from its neighbors at time $t$ is $R_{i}^{\pi}(t)=\sum_{j\in\mathcal{N}{}_{i}}D_{ji}^{\pi}(t)$ and the average amount of resource received by time $t$ is, 
$\bar{R}_{i}^{\pi}\left(t\right)=\sum_{\tau=1}^{t}R_{i}^{\pi}\left(\tau\right)/t$. We denote the long term average resource that user $i$ receives under policy $\pi$ as $\liminf_{t\rightarrow\infty}\bar{R}_{i}^{\pi}\left(t\right)\triangleq r_{i}^{\pi}$.
Note that in general $r_{i}^{\pi}$ is a random variable. However, as we will see in the next section, in order to obtain policies that satisfy the objectives of interest in this work, it suffices to restrict attention to policies for which $\lim_{t\rightarrow\infty}\bar{R}_{i}^{\pi}\left(t\right)$ exists and has a constant and finite value. 

The set of feasible long term average received resource vectors that can be achieved by policies in $\Pi$ is denoted by $\mathcal{R}$. That is, 
\begin{equation}
\mathcal{R}=\big\{\bm{r}^{\pi}=(r_{i}^{\pi})_{i\in\mathcal{N}}:\ \pi\in\Pi\big\}.\label{eq:RateSpace}
\end{equation}
Node $i$ is interested in maximizing its long-term average received resource $r_{i}^{\pi}$. Clearly, the objectives of nodes are conflicting, as neighbors have to compete for the same resources and therefore the key issue is to decide how to allocate the resources produced by the nodes. There are two basic approaches to address this issue. Namely, one could formulate this problem as a centrally defined fair-allocation problem and take into account the resource contribution of each node $i\in\mathcal{{N}}$ to the community in the long-term, so as to decide how much resource $r_{i}$ to return to it. In a different context, each node is interested in maximizing it own received resource,  and this gives rise to competitive interactions and hence creates a sharing economy market. In that case, the amount of resource each node receives in the long run depends on the attained equilibrium, if any exist. Additionally, it is possible in some settings that users can coordinate with each other and form sharing groups, or coalitions, aspiring to improve their benefits by excluding non-members. Our goal is to analyze the long-term average performance of the nodes' dynamic interactions in the three frameworks described above. This is formalized in the next subsection.
\begin{figure}[t]
	\centering
	\includegraphics[scale=0.33]{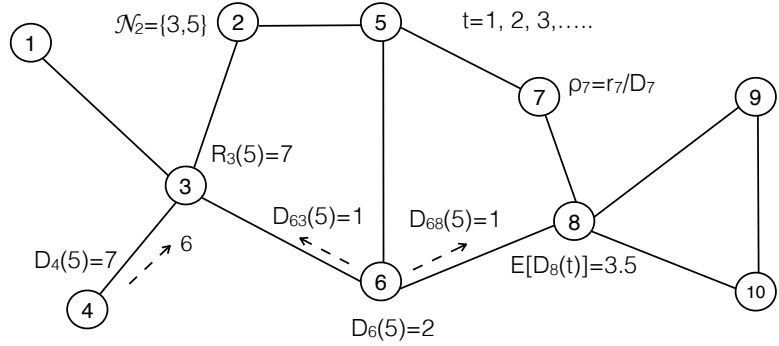}\caption{{\small{}{The model of a sharing economy network. The different system parameters and node decisions are presented, for the long-term regime or a specific time slot $t=5$.}\label{fig:system-model}}}
\end{figure}

\subsection{\label{subsec:Operating-Frameworks}Operating Frameworks}

In this section, we consider three different problem formulations whose objectives are based on long-term averages of the quantities of interest.

\subsubsection{Fairness Framework}

In this setting, we consider a centralized policy designed to allocate resources to nodes in proportion to their contribution to the community. Ideally, in such a setting each nodes distributes in the long-run all its endowment and it would be desirable to allocate to every node $i\in\mathcal{N}$ resource equal to its contribution, i.e., $r_{i}=D_{i}$. However, due to the resource sharing constraints imposed by the graph and the different resource endowments of the nodes, such policies will not be feasible in general. Given this, the designer would prefer to ensure the ``most balanced'' long-term allocation. A suitable method to achieve this goal is to employ the lexicographic optimal (or, lex-optimal) criterion, which has been extensively used for resource allocation and load balancing, for example, in communication networks \citep{georgiadislexopt02}, \citep{nace-tutorial}, \citep{boudecfairness07}. This multi-objective optimization method first increases as much as possible the allocated resource to the node with the smaller sharing ratio, $r_{i}/D_{i}$. Next, if there are many choices, it attempts to increase the resource allocated to the node with the second smaller sharing ratio, and so on. The resulting long-term average allocation is max-min fair, thus as balanced as possible. Next we provide the necessary definitions.
\begin{definition}
\textbf{Lexicographical order}. Let $\bm{x}$ and $\bm{y}$ be $N$-dimensional vectors, and $\boldsymbol{\phi}(\bm{x})$ and $\boldsymbol{\phi}(\bm{y})$	the $N$-dimensional vectors that are created by sorting the components	of $\bm{x}$ and $\bm{y}$ respectively, in non-decreasing order.	We say that $\bm{x}$ is lexicographically larger than $\bm{y}$,	denoted by $\bm{x}\succ\bm{y}$, if the first non-zero component of	the vector $\phi(\bm{x})-\phi(\bm{y})$ is positive. The notation	$\bm{x}\succeq\bm{y}$ means that either $\bm{x}\succ\bm{y}$ or,	$\bm{x}=\bm{y}$. 
\end{definition}

Within this framework we are interested determining policies that induce lexicographically optimal \emph{sharing ratio} vector, i.e.,
defining $\rho_{i}^{\pi}\triangleq\frac{r_{i}^{\pi}}{D_{i}},\,\, \boldsymbol{\rho}^{\pi}=\left(\rho_{i}^{\pi}\right)_{i\in\mathcal{N}}$, we are interested in determining a policy $\pi^{*}$ such that $\boldsymbol{\rho}^{\pi^{*}}\succeq\boldsymbol{\rho}^{\pi},\ \text{for all \ensuremath{\pi\in\Pi}}$. In the following, a vector $\boldsymbol{r}$ whose sharing ratio vector is lexicographically optimal will also be called simply ``lexicographically optimal'' or ``lex-optimal''.

\subsubsection{Competitive Framework}

Assume now that each node $i\in\mathcal{N}$ is an independent decision maker, interested in maximizing the long-term average resource $r_{i}$ it receives. An approach in this setup is to define exchange ratios for the node resources that have the following property: each node receives resources from its neighbors in such a manner that the node maximizes its received resource subject only to the constraint that the cost of received resource does not exceed its wealth determined by the exchange ratio and the size of its endowment (the constraints regrading the size of the endowments of the neighbors of the node are not taken into account in this optimization). The solution concept for this setup is effectively the competitive (or, Walrasian) equilibrium \citep{arrow-debreu}, \citep{ColellWhinstonGreenBook1995}, which has been also applied in communication networks \citep{RJohariToNBilateral2011}, and extended to graphical economies (which exhibit \emph{localities})
\citep{KearnsGraphEcon2004,KearnsEconSocial2004}. However, for the problem under consideration, we avoid explicit exchange ratios, and introduce a closely related equilibrium concept: 
\begin{definition}
\textbf{\emph{Sharing Equilibrium.}} A sharing equilibrium is determined by a vector of sharing ratios $\boldsymbol{\rho}^{*}=\left(\rho_{i}^{*}\right)_{i\in\mathcal{N}}$ with the following properties. a) If at time $t$ node $i$ gives resource $D_{ij}(t)$ to node $j\in\mathcal{N}_{i}$, node $i$ expects in return (either at time $t$ or in the future) resource\footnote{This return resource may be obtained either from node $j$ or from any other neighbor of node $i$. This can be interpeted as follows. If node $i$ provides resource $D_{ij}(t)$ to node $j,$ it gets resource credit $D_{ij}(t)/\rho_{j}^{*}$. Node $i$ can receive this amount of resource from any of its neighbors either at time $t$ or at some time in the future. } $D_{ij}(t)/\rho_{j}^{*}$, and b) node $i$ gives the resources it generates to its neighbors in such a manner that it maximizes its received resource in the long run, $r_{i},$ under the constraint that it does not exceed the amount it is entitled by the specified exhange rate and its endowment, that is, $r_{i}\leq D_{i}\rho_{i}^{*}$. It is easy to see that the resulting policy $\pi^{*}$ should be such that,
\end{definition}
\begin{enumerate}
	\item \label{enu:exhEq1}each node distributes all its endowment to its
	neighbors in the long run, i.e., 
	\begin{equation}
	\lim_{t\rightarrow\infty}\bar{D}_{i}\left(t\right)=D_{i},\label{eq:exchange0}
	\end{equation}
	\item \label{enu:exEq2}each node distributes its generated resource at
	all times to the neighbors that have the smallest sharing ratio. Moreover,
	since each node attempts to maximize its received resource without
	taking into account the avaibable endowments of its neighbors, the
	optimization should result in received resource rate vector $\boldsymbol{r}^{*}$
	that satisfies: $r_{i}^{*}=D_{i}\rho_{i}^{*},\ i\in\mathcal{N}$.\label{eq:exchange}
\end{enumerate}

In this context, we are interested in determining whether equilibrium sharing ratios and associated policies exist. Moreover, we are interested in determining policies that operate without a priori knowledge of the equilibrium rates (provided that they exist), but adjust sharing ratios over time in such a manner that they eventually converge to the equilibrium ones; in addition, the long-term received resources are those that are obtained by employing policies that know a priori the equilibrium sharing ratios.

\subsubsection{Coalitional Framework}

Before providing the details of this framework, let us introduce some additional notation. We denote by $G_{\mathcal{S}}=\left(\mathcal{S},\mathcal{E_{\mathcal{S}}}\right)$ the subgraph of $G$ induced by a nonempty set of nodes $\mathcal{S}\subseteq\mathcal{N}$, i.e., the graph with node set $\mathcal{{S}}$, and link set $\mathcal{E}_{\mathcal{S}}=\left\{ \left(i,j\right):\ i,j\in\mathcal{S}\right\}$. We denote by $\Pi_{\mathcal{S}}$ the set of policies that operate on graph $G_{\mathcal{S}}$ and by $\mathcal{R}_{\mathcal{{S}}}$ the set of all long-term received resource vectors that can be obtained by employing policies in $\Pi_{\mathcal{S}},$
\[
\mathcal{R}_{\mathcal{S}}=\big\{\bm{r}^{\pi}=(r_{i}^{\pi})_{i\in S}:\ \pi\in\Pi_{\mathcal{S}}\big\}.
\]
Note that the Graph $G_{\mathcal{S}}$ may not be connected. However, the definition of policy in Section 2.1 still holds an hence the set of policies $\Pi_{\mathcal{S}}$ is well defined. Also, all the stated results for connected graphs hold for each of the connected components of $G_{\mathcal{S}}$.

In this setting we assume that subsets of nodes can coordinate to form coalitions and deviate from the proposed fair solution if this will ensure higher resources for some of them. In game theoretic terms, this behavior leads to a coalitional (or, cooperative) game \citep{myerson-gametheory-book} played by the nodes. Specifically, we call any nonempty subset of nodes $\mathcal{{S}}\subseteq\mathcal{{N}}$ a coalition when they allocate their resources only among each other. That is, there is no resource exchange among nodes in $\mathcal{S}$ and nodes in its complement set $\mathcal{S}^{c}=\mathcal{N}-\mathcal{S}$. Hence,
the feasible long-term resource vectors that nodes in $\mathcal{{S}}$ get, are the $\left|\mathcal{S}\right|-$dimensional vectors in $\mathcal{R}_{\mathcal{S}}$. We refer to the set $\mathcal{N}$ as the grand coalition. This coalitional game is one with non-transferable utilities, as resources cannot be split arbitrary among the nodes, due to the exchange constraints imposed by the graph. Our goal is to study the existence and the properties of self-enforcing long-term allocations. This property is formally captured by the notion of stability for the grand coalition. 
\begin{definition}
\textbf{Coalitional Stability}. A grand coalition $\mathcal{N}$ along with a policy $\pi^{*}\in\Pi$ that induces long-term received resource vector $\bm{r}^{*}$ is called \emph{strongly} stable if for any nonempty node set $\mathcal{{S}}\subseteq\mathcal{{N}}$ , there is no policy $\pi_{\mathcal{S}}\in\Pi_{\mathcal{S}}$ that induces an ($\left|\mathcal{S}\right|-$dimensional)	vector $\boldsymbol{r}$ such that $r_{i}\geq r_{i}^{*}$ for all $i\in\mathcal{{S}}$, and $r_{j}>r_{j}^{*}$ for at least one node	$j\in\mathcal{{S}}$. The allocation is called \emph{weakly} stable if for any nonempty node set $\mathcal{{S}}\subseteq\mathcal{{N}}$, there is no policy $\pi_{\mathcal{S}}\in\Pi_{\mathcal{S}}$ that induces a vector $\boldsymbol{r}^{\mathcal{S}}$ such that $r_{i}^{\mathcal{S}}>r_{i}^{*}$ for all $i\in\mathcal{{S}}$.
\end{definition}
Note that strong stability implies weak stability but not the other way around. In particular, the concept of weak stability for the grand coalition is directly related to the concept of the \emph{core}. In this coalitional framework, we ask the question: Is there a policy $\pi\in\Pi$ that renders the grand coalition stable?

\section{A Unifying Policy For The Three Frameworks } \label{sec:algorithms}

In this section we describe a simple policy $\pi^{*}$ that achieves the objectives of all three frameworks defined in Section \ref{subsec:Operating-Frameworks}. According to $\pi^{*}$ each node maintains a ratio $\rho_{i}\left(t\right)=\bar{R}_{i}\left(t\right)/D_{i}$ which may be interpreted as resource sharing ratio (or simply ``ratio'')
at time $t.$ At time $t,$ every node gives its generated resource to the node that has the smallest sharing ratio among its outgoing neighbors. Specifically, the policy operates according to Algorithm \ref{alg:1}. Note that the only a priori information required for the operation of the policy is the set of endowments of the nodes. However, as will be discussed in Section \ref{sec:conclusions}, the policy can also operate by replacing $D_{i}$ with the time average $\bar{D}_{i}\left(t\right)=\sum_{\tau=1}^{t}D_{i}\left(\tau\right)/t$.

\begin{algorithm}[h]

\nl At time $t=1$ set $\rho_{i}\left(t\right)=0,\ i\in{\mathcal{N}}$
\\
\nl At the beginning of slot $t$, 

\nl Each user $i\in{\mathcal{N}}$ announces to its neighbors
the sharing ratio $\rho_{i}(t)=\bar{R}_{i}(t)/D_{i}$, 

\nl Each user $i\in{\mathcal{N}}$ distributes the resource it generates at time $t$ to its neighbor(s) $j\in\mathcal{N}_{i}$ that have the smallest sharing ratio, $\rho_{j}(t)$. 

\caption{\label{alg:1} Algorithm according to which policy $\pi^{*}$ operates}

\end{algorithm}

The next theorem is the main result of this work. 
\begin{theorem}
\label{thm:Optimality-1}The following hold.
\begin{itemize}
\item Policy $\pi^{*}$ is Lexicographically optimal. 
\item Under $\pi^{*}$the node sharing ratios and long-term received resources converge to the equilibrium sharing ratios and equilibrium received resources. 
\item Policy $\pi^{*}$ is coalitionally stable.
\end{itemize}
\end{theorem}
Next we present an outline of the arguments that will be used to prove this theorem. In Section \ref{subsec:AchievableRates} we show that the region $\mathcal{R}$ is a subset of a polymatroid, and this allows us to restrict attention to policies under which all nodes distribute their endowments to their neighbors in the long run. In Section \ref{subsec:Structure}, using the structure of the lexicographically optimal vector in polymatroids, we derive the structure for the lexicographically optimal point in our setup, and we use this structure to show that a policy that achieves the lexicographically optimal received resource vector also achieves
the equilibrium sharing ratios in the long run. Also, we show that a policy that achieves the equilibrium sharing ratios is stable. Finally, in the Appendix we show that policy $\pi^{*}$ achieves the lexicographically optimal point and thus possesses all properties described in Theorem \ref{thm:Optimality-1}.

\subsection{\label{subsec:AchievableRates}Achievable Received Resource Vectors}

In this section we provide some important properties about the region $\mathcal{R}$ in (\ref{eq:RateSpace}) which consists of the possible vectors of long-term average received resources $\boldsymbol{r}^{\pi}=\left(r_{i}^{\pi}\right)_{i=1}^{N}$ that can be obtained by any policy $\pi\in\Pi$. For a set $\mathcal{{S\subseteq\mathcal{{N}}}}$ define by $\mathcal{{N}}_{\mathcal{{S}}}$ the set of nodes that are neighbors of nodes in $\mathcal{{S}}$, i.e.,  $\mathcal{{N}}_{\mathcal{{S}}}=\cup_{i\in\mathcal{{S}}}\mathcal{{N}}_{i},\,\ {\mathcal{N}}_{\emptyset}=\emptyset.$
Let 
\[
f(\mathcal{{S}})=\sum_{i\in\mathcal{{N}}_{\mathcal{{S}}}}D_{i}
\]
where for $\mathcal{G}=\emptyset$ we define $\sum_{i\in\mathcal{G}}x_{i}=0$. Since there are no isolated nodes, $f\left(\mathcal{N}\right)=\sum_{i\in\mathcal{N}}D_{i}$. The following lemma is a consequence of the fact that nodes can receive
resources only from their neighbors. 
\begin{lemma}
\label{lem:inequality-1}Under any policy $\pi\in\Pi$ it holds for
any $\mathcal{{S}}\subseteq\mathcal{{N}},$ 
\begin{align}
\limsup_{t\rightarrow\infty}\sum_{i\in{\mathcal{S}}}\bar{R}_{i}^{\pi}\left(t\right) & \leq f({\mathcal{S}}),\label{eq:basic1-2}\\
\sum_{i\in{\mathcal{S}}}r_{i}^{\pi} & \leq f\left({\mathcal{S}}\right).\label{eq:basic0}
\end{align}
\end{lemma}

Let $\Pi_{0}$ be the class of policies in $\Pi$ for which 
\begin{enumerate}
\item The long-term average of received resources exist, i.e, $r_{i}^{\pi}=\lim_{t\rightarrow\infty}\sum_{\tau=1}^{t}{R_{i}^{\pi}\left(\tau\right)}/{t},\ i\in{\mathcal{N}}$. 
\item All endowments generated by the nodes are eventually consumed, i.e.,
$\sum_{i\in\mathcal{{N}}}r_{i}^{\pi}=f\left(\mathcal{N}\right).$ 
\end{enumerate}
Let ${\mathcal{R}}_{0}$ be the set of received resource vectors, $\boldsymbol{r},$ that can be achieved by policies in $\Pi_{0}$. From Lemma \ref{lem:inequality-1} we conclude that 
\begin{align}
{\mathcal{\mathcal{R}}} & \subseteq\left\{ \boldsymbol{r}\geq\boldsymbol{0}:\,\sum_{i\in{\mathcal{S}}}r_{i}\leq f({\mathcal{S}}),\ {\mathcal{S}}\subseteq{\mathcal{N}}\right\} \triangleq\mathcal{A},\label{eq:subset}\\
\mathcal{R}_{0} & \subseteq\left\{ \boldsymbol{r}\geq\boldsymbol{0}:\,\sum_{i\in{\mathcal{S}}}r_{i}\leq f({\mathcal{S}}),\ {\mathcal{S}}\subset{\mathcal{N}},\ \sum_{i\in{\mathcal{N}}}r_{i}=f({\mathcal{N}})\right\} \triangleq\mathcal{A}_{0}.\label{eq:subset0}
\end{align}
To proceed, we need the fact that $f\left(\mathcal{S}\right)$ possesses the important property of submodularity.
\begin{lemma}
\label{lem:f is submodular.-1}$f(\mathcal{{S}})$ is submodular i.e.,
it holds for every $\mathcal{S},\ \mathcal{T}\subseteq\mathcal{N},$
\begin{equation}
f({\mathcal{S}}\cap{\mathcal{T}})+f\left({\mathcal{S}}\cup{\mathcal{T}}\right)\leq f({\mathcal{S}})+f({\mathcal{T}}).\label{eq:submodar-1}
\end{equation}
\end{lemma}
For submodular $f\left(\mathcal{S}\right)$, the sets $\mathcal{A}$ and $\mathcal{A}_{0}$ are referred to as ``polymatroid polyhedron'' and ``base of the polymatroid'' respectively. Using the polymatroid property, the next lemma shows that the achievable resource vectors under policies in $\Pi_{0}$ is the base of the polymatroid. 
\begin{lemma}
\label{lem:polybase}It holds: $\mathcal{R}_{0}=\mathcal{A}_{0}.$
\end{lemma}

\subsection{Review of Polymatroid Properties}

In this section we present some properties of polymatroids that are needed in the sequel \citep{submodular-book, tsoucas-MOR}.
\begin{lemma}
\label{lem:reduce}If $\mathcal{A}$ is a polymatoid with base $\mathcal{A}_{0}$, then for any $\boldsymbol{r}\in\mathcal{A}$ there exist an $\boldsymbol{r}_{0}\in\mathcal{A}_{0}$ such that $\boldsymbol{r}_{0}\leq\boldsymbol{r}$. Hence the lexicographically
optimal vector in $\mathcal{A}$ lies in $\mathcal{A}_{0}.$
\end{lemma}
Next we describe the structure of the lexicographically optimal vector in $\mathcal{A}$. First we need some notation. For a given $\bm{r}\in\mathcal{A}$, the \emph{different values} the coordinates of vector $\bm{\rho}=\left(\rho_{i}\right)_{i\in\mathcal{N}}=\left(r_{i}/D_{i}\right)_{i\in\mathcal{N}}$ takes, will be denoted by $v_{k}\left(\boldsymbol{r}\right),\ i=1,...K(\boldsymbol{r)}\leq N$, where $v_{1}\left(\boldsymbol{r}\right)<v_{2}\left(\boldsymbol{r}\right)<...<v_{K(\boldsymbol{r})}\left(\boldsymbol{r}\right)$. The index of the value to which $\rho_{i}$ equals is denoted by $I_{i}\left(\boldsymbol{r}\right)$, i.e., $v_{I_{i}\left(\boldsymbol{r}\right)}=\rho_{i}$. We call $I_{i}\left(\boldsymbol{r}\right)$ the ``level of node $i$''. The set of nodes of level $k$ is denoted by $\mathcal{L}_{k}\left(\boldsymbol{r}\right)=\left\{ i\in\mathcal{N}:\ I_{i}\left(\boldsymbol{r}\right)=k\right\} $. 
\begin{theorem}
\label{thm:lextoptvector}Let $\mathcal{A}$ be a polymatroid. A vector $\boldsymbol{r}$ in $\mathcal{A}$, is lexicographically optimal if and only if the following hold.
\begin{align}
\sum_{i\in\mathcal{L}_{1}}r_{i} & =f\left(\mathcal{L}_{1}\right),\label{eq:lex1}\\
\sum_{i\in\mathcal{L}_{k}}r_{i} & =f\left(\cup_{l=1}^{k}\mathcal{L}_{l}\right)-f\left(\cup_{l=1}^{k-1}\mathcal{L}_{l}\right),\ 2\leq k\leq K,\label{eq:lex2}
\end{align}
where, $\mathcal{L}_{k}=\mathcal{L}_{k}\left(\boldsymbol{r}\right).$ $K=K\left(\boldsymbol{r}\right).$ The lexicographically optimal vector exists and is unique.
\end{theorem}

\subsection{\label{subsec:Structure}Structure of Lexicographically Optimal Received Resource Vector }

\begin{figure}
\centering
\includegraphics[scale=0.3]{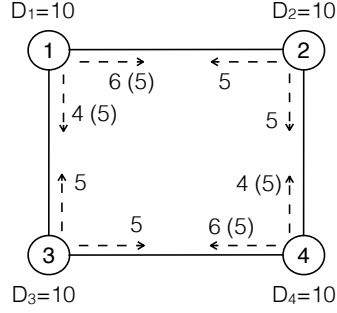}
\caption{Example of Multiple Allocations leading to identical sharing ratios. Numbers show main allocation rates, while in parentheses the alternative options are depicted. All nodes have the same average resource generation rate. Both allocations lead to the same exchage ratio vector of (1,1,1,1).}
\label{fig:example-with-2-allocations}
\end{figure}

In this section we describe the structure of the lexicographically optimal vector for the problem under consideration. We will make use of the following simple lemma.
\begin{lemma}
\label{lem:levels}Let $\boldsymbol{r}\in\mathcal{A}_{0}$ then: a) If $K=1$ then $v_{1}=1$. b) If $K>1$ , then $v_{1}<1$ and $l_{K}>1$.
\end{lemma}
%
%
For the problem under consideration in this work it can be easily seen that given any vector $\boldsymbol{r}$ in $\mathcal{A}_{0}$
there is an allocation set $\left\{ d_{ij}\geq0,\ i\in\mathcal{N},\ j\in\mathcal{N}_{i}\right\} ,$
such that $\sum_{j\in\mathcal{N}_{i}}d_{ij}=D_{i},\ i\in\mathcal{N}\label{eq:allocgive}$ and 
$\sum_{j\in\mathcal{N}_{i}}d_{ji}=r_{i},\ i\in\mathcal{N}\label{eq:alloctake}$, we refer to this set as ``allocation that generates $\boldsymbol{r}.$'' This allocation may not be unique as can be seen in the example shown in Figure \ref{fig:example-with-2-allocations}. Fixing any such allocation, we say that (under this allocation) ``node $i$ gives resource to node $j$'' if $d_{ij}>0$. We also say that ``node $i$ gives resource to a set $\mathcal{S}$'', if node $i$ gives resource to any node
in $\mathcal{S}.$

Consider now the general structure of the lexicographically optimal vector described in Theorem \ref{thm:lextoptvector}. For the problem under consideration in the current work, equality (\ref{eq:lex1}) implies that there is (at least one) allocation set $\left\{ d_{ij}\geq0,\ i\in\mathcal{N},\ j\in\mathcal{N}_{i}\right\} ,$ such that the endowments of all neighbors of set $\mathcal{L}_{1}$ are given to the nodes in this set. Similarly, (\ref{eq:lex2}) for $k=2$ implies that the endowments of all nodes in $\mathcal{N}_{\mathcal{L}_{2}}-\mathcal{N}_{\mathcal{L}_{1}}$are given to the nodes in set $\mathcal{L}_{2}$; in general, the endowments of all nodes in $\mathcal{N}_{\mathcal{L}_{k}}-\mathcal{N}_{\cup_{l=1}^{k-1}\mathcal{L}_{l}}$ are given to nodes in set $\mathcal{L}_{k}.$ To proceed we need some additional notation. Given vector $\boldsymbol{r}$ with $K\left(\boldsymbol{r}\right)\geq2$ define,
\begin{align}
\mathcal{Q}_{k}  \left(\boldsymbol{r}\right)=\mathcal{N}-\cup_{m=1}^{k-1}\left(\mathcal{L}_{m}\left(\boldsymbol{r}\right)\cup\mathcal{L}_{K-m+1}\left(\boldsymbol{r}\right)\right),\ 1\leq k\leq\lceil{K/2\rceil},
\end{align}
where $\cup_{m=k}^{l}\mathcal{S}=\emptyset$ if $l<k.$ For example, $\ensuremath{\mathcal{Q}_{1}(\boldsymbol{r})=\mathcal{N}}$, and $\mathcal{Q}_{2}\left(\boldsymbol{r}\right)$ consists of the nodes in $\mathcal{N}$ that remain after removing those that belong to level sets $\mathcal{L}_{1}\left(\boldsymbol{r}\right)$ and $\mathcal{L}_{K}\left(\boldsymbol{r}\right)$. In the sequel,
a quantity $X$ referring to induced subgraph $G_{\mathcal{Q}_{k}\left(\boldsymbol{r}\right)}=\left(\mathcal{Q}_{k}\left(\boldsymbol{r}\right),\mathcal{E}_{\mathcal{Q}_{k}}\left(\boldsymbol{r}\right)\right)$ is denoted $X_{\mathcal{Q}_{k}\left(\boldsymbol{r}\right)}$. Also when there is no possibility for confusion, for simplicity we omit from the notation the dependence of the quantities defined above on the vector $\boldsymbol{r}.$

The next Theorem describes the structure of the lexicographically optimal vector. 
\begin{theorem}
\label{thm:MainTh0-1} A vector $\boldsymbol{r}\in\mathcal{\mathcal{A}}$, is lexicographically optimal if and only if the following hold. If $K=1,$ then $v_{1}=1$. If $K\geq2$ then
\begin{enumerate}
\item \label{enu:MainTh0Item1-1}$\mathcal{L}_{k}$ is an independent set in graph $G_{\mathcal{Q}_{k}}$, for $k=1,....,\lfloor{\frac{K}{2}\rfloor}$. 
\item \label{enu:MainTh0Item2-1}$\mathcal{L}_{K-k+1}=\mathcal{N}_{\mathcal{Q}_{k}}\left(\mathcal{L}_{k}\right)$, for $k=1,....,\lfloor{\frac{K}{2}\rfloor}$. 
\item \label{enu:MainTh0Item2.1-1}$v_{k}v_{K-k+1}=1$, for $k=1,....,\lfloor{K/2\rfloor}$. 
\item \label{enu:MainTh0Item3-1}$\sum_{i\in\mathcal{L}_{k}}r_{i}=\sum_{i\in\mathcal{L}_{K-k+1}}D_{i}$, for $k=1,....,\lfloor{\frac{K}{2}\rfloor}$. 
\item \label{enu:main-5}If $K$ is odd, then $v_{\left\lceil K/2\right\rceil }=1.$
\end{enumerate}
\end{theorem}
\noindent 
\begin{figure}[t]
\begin{centering}
\includegraphics[scale=0.4]{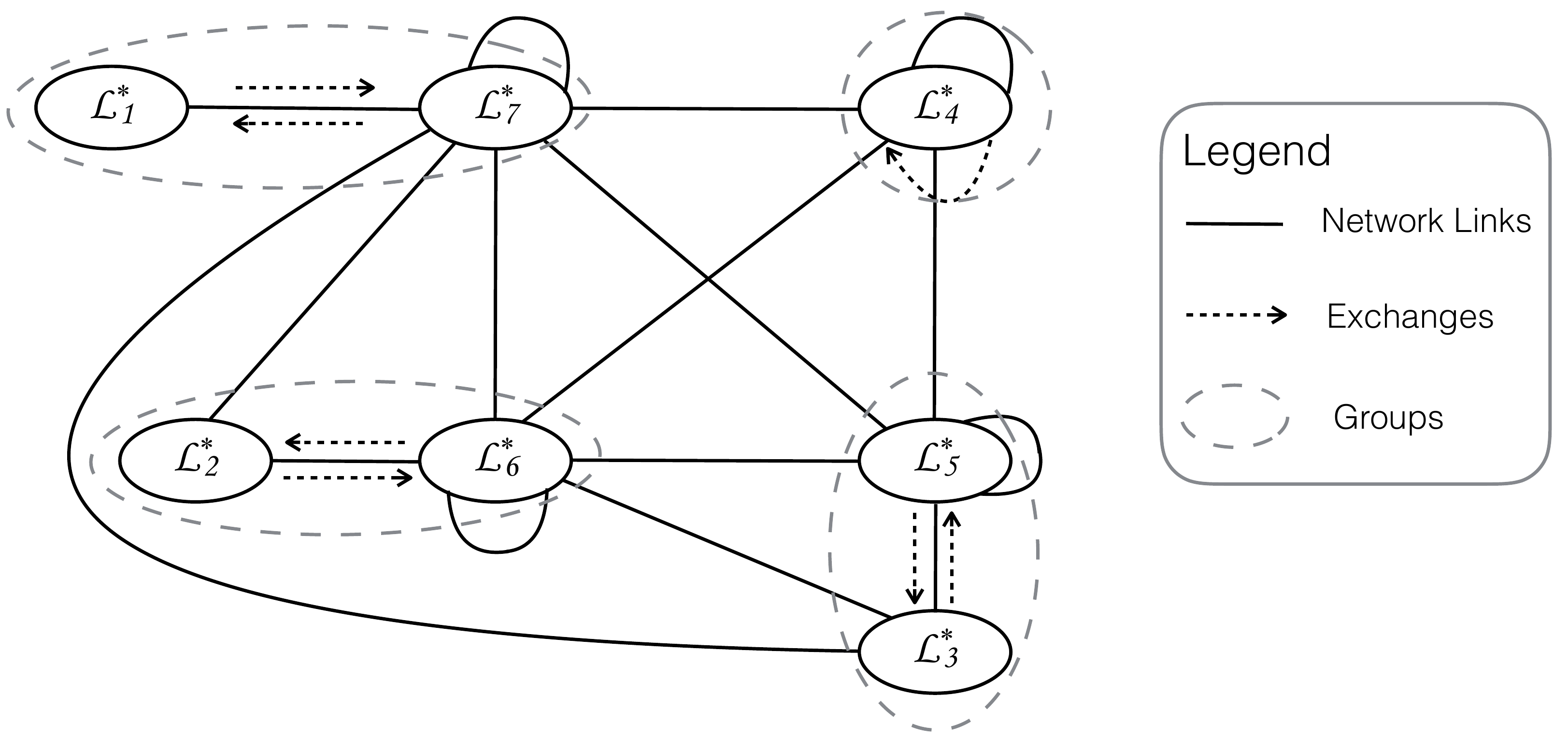}
\par\end{centering}
\caption{{Structure of a graph with $K^{*}=7$ levels. Solid lines show the possible exchanges, or, the physical connections, while dashed arrows the actual exchanges. Note that not all connections are employed, as some groups of players do not interact (or, collaborate) with each other.}\label{fig:Corollary-Example}}
\end{figure}
Let us now discuss the implications of the above theorems. Under a lexicographically optimal allocation $\boldsymbol{r}^{*}$, the nodes are divided in disjoint sets $\mathcal{L}_{1}^{*},\cdots,\mathcal{L}_{K^{*}}^{*}$, where nodes in each set have the same sharing ratio; this division depends both on the resource endowments of the nodes and on the graph $G$. For the discussion below, please refer to Fig. \ref{fig:Corollary-Example}, that presents an example of the structure for $K^{*}=7$ levels. In this graph, we depict with solid lines the physical connections that may exist among the different sets of nodes. Notice that the actual nodes and their detailed connections are not shown. 

\underline{Sharing ratios Structure}. According to Theorem \ref{thm:MainTh0-1}, the sharing ratios have a certain structure. Specifically, the highest sharing ratio is inversely proportional to the lowest one ($l_{7}^{*}=1/l_{1}^{*}$), the second highest sharing ratio is inversely proportional to the second lowest sharing ratio ($l_{6}^{*}=1/l_{2}^{*}$), and so on. Additionally, as shown in Fig. \ref{fig:Corollary-Example} all the nodes in the set with the highest sharing ratio, $\mathcal{L}_{7}^{*}$, exchange resources only with nodes belonging to the set with the lowest sharing ratio, $\mathcal{L}_{1}^{*}$. Similarly, the nodes in set
$\mathcal{L}_{6}^{*}$ exchange resources only with nodes in set $\mathcal{L}_{2}^{*}$ and so on. When $K^{*}$ is odd, there is one set of nodes, here the set $\mathcal{L}_{4}^{*}$, which exchange resource only with each other.

\underline{Topological Properties}. Nodes in the set with the lowest exchange rate, $\mathcal{L}_{1}^{*}$ constitute an independent set. Moreover, the set of their neighbors is the set with the highest sharing ratio, $\mathcal{L}_{K^{*}}^{*}.$ Similarly, it holds that $\mathcal{L}_{6}^{*}=\mathcal{N}_{\mathcal{Q}_{2}}(\mathcal{L}_{2}^{*})$ nodes in $\mathcal{L}_{2}^{*}$ constitute an independent set in the graph $G_{\mathcal{Q}_{2}\left(\boldsymbol{r^{*}}\right)}$. Hence, the nodes in set $\mathcal{L}_{2}^{*}$ can have links only with nodes in set $\mathcal{L}_{6}^{*}$ and possibly with nodes in $\mathcal{L}_{7}^{*}$ (since the latter do not belong in $G_{\mathcal{Q}_{2}\left(\boldsymbol{r}^{*}\right)}$). However, as discussed above, nodes in $\mathcal{L}_{2}^{*}$ exchange resource only with nodes in $\mathcal{L}_{6}^{*}$. With the same reasoning, it is easy to see that nodes in set $\mathcal{L}_{3}^{*}$ can be physically connected with nodes in $\mathcal{L}_{7}^{*}$, $\mathcal{L}_{6}^{*}$ and $\mathcal{L}_{5}^{*}$, but they exchange resource only with nodes in the latter set. Finally, nodes in set $\mathcal{L}_{4}^{*}$ exchange resources only with each other.

These properties reveal how the graph affects the lex-optimal fair solution. For example, by adding a link between two nodes initially belonging to $\mathcal{L}_{1}^{*}$ (which is independent), the lex-optimal solution changes and places these (now connected) nodes to another set. This dependency among the graph structure and the lex-optimal sharing ratio vector will become more evident in the sequel.

The next theorem shows that the ratios of the lexicographically optimal point are the equilibrium sharing ratios of the competitive framework. 
\begin{theorem}
\label{thm:competitive}Let $\boldsymbol{r}^{*}$ be a lexicographically optimal vector. The ratios $\left\{ \rho_{i}^{*}\right\} _{i\in\mathcal{N}}=\left\{ r_{i}^{*}/D_{i}\right\} _{i\in\mathcal{N}}$ are equilibrium sharing ratios for the competitive framework. 
\end{theorem}

The next theorem shows that a policy achieving the lexicographically optimal vector $\boldsymbol{r}^{*}$ is stable. 
\begin{theorem}
\label{thm:stable}A policy $\pi^{*}$ that achieves the lexicographically optimal vector $\boldsymbol{r}^{*}$is strongly stable.  
\end{theorem}

The next section presents representative numerical examples that shed light on the above results.

\section{Numerical Results and Discussion}\label{sec:Numerical-Results}

For our numerical investigation, we consider two families of network graphs. First, we focus on small graphs with typical structures, such as rings. Then, we present the equilibriums in well-known network models such as Erdos-Renyi, Lattice, Scale-free and Small-world graphs.

\subsection{Basic Graphs}

\begin{figure}[t]
\begin{subfigure}[b]{0.23\textwidth}	
\includegraphics[scale=0.2]{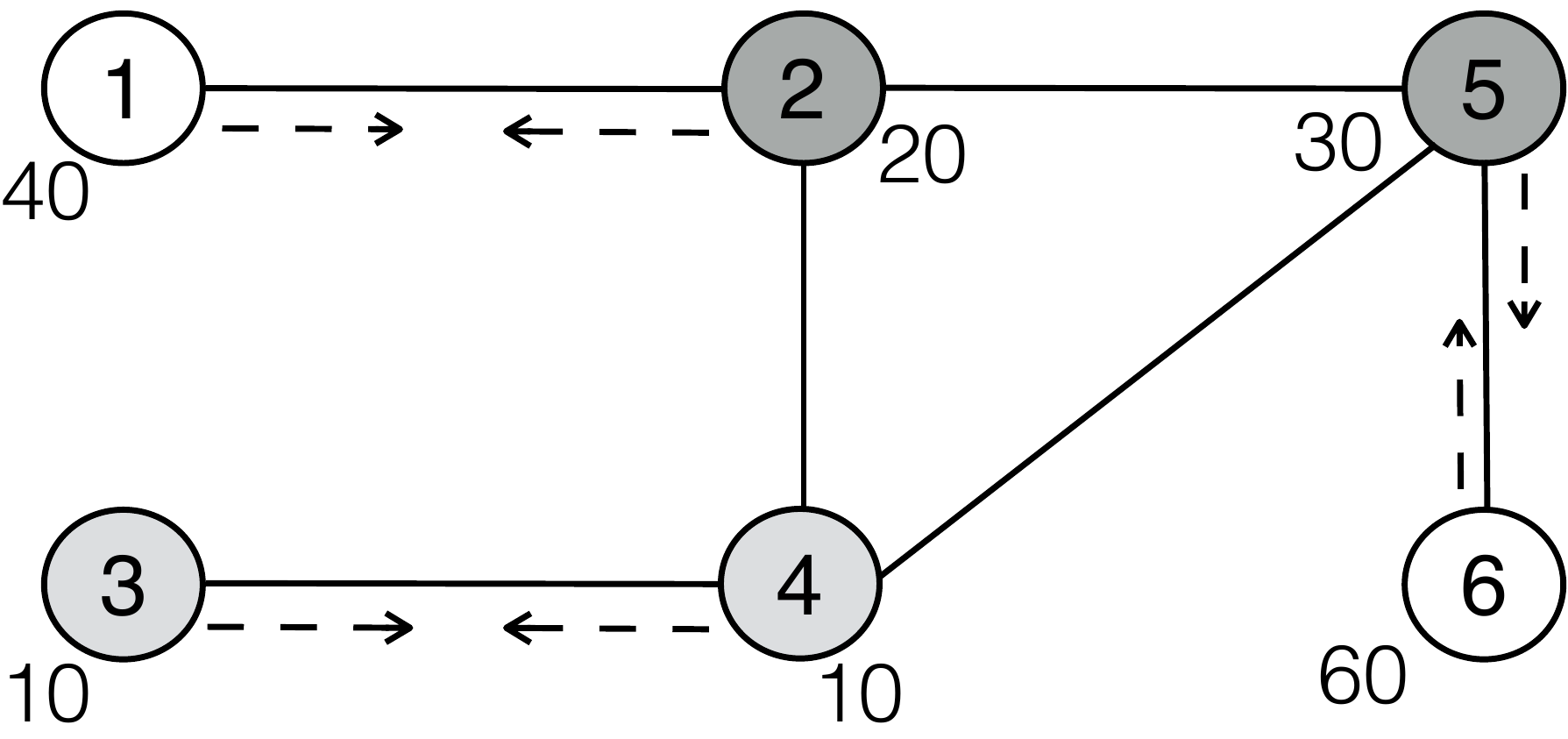}
\caption{}
\label{fig:simple-example-1}
\end{subfigure}
\begin{subfigure}[b]{0.23\textwidth}	
\includegraphics[scale=0.2]{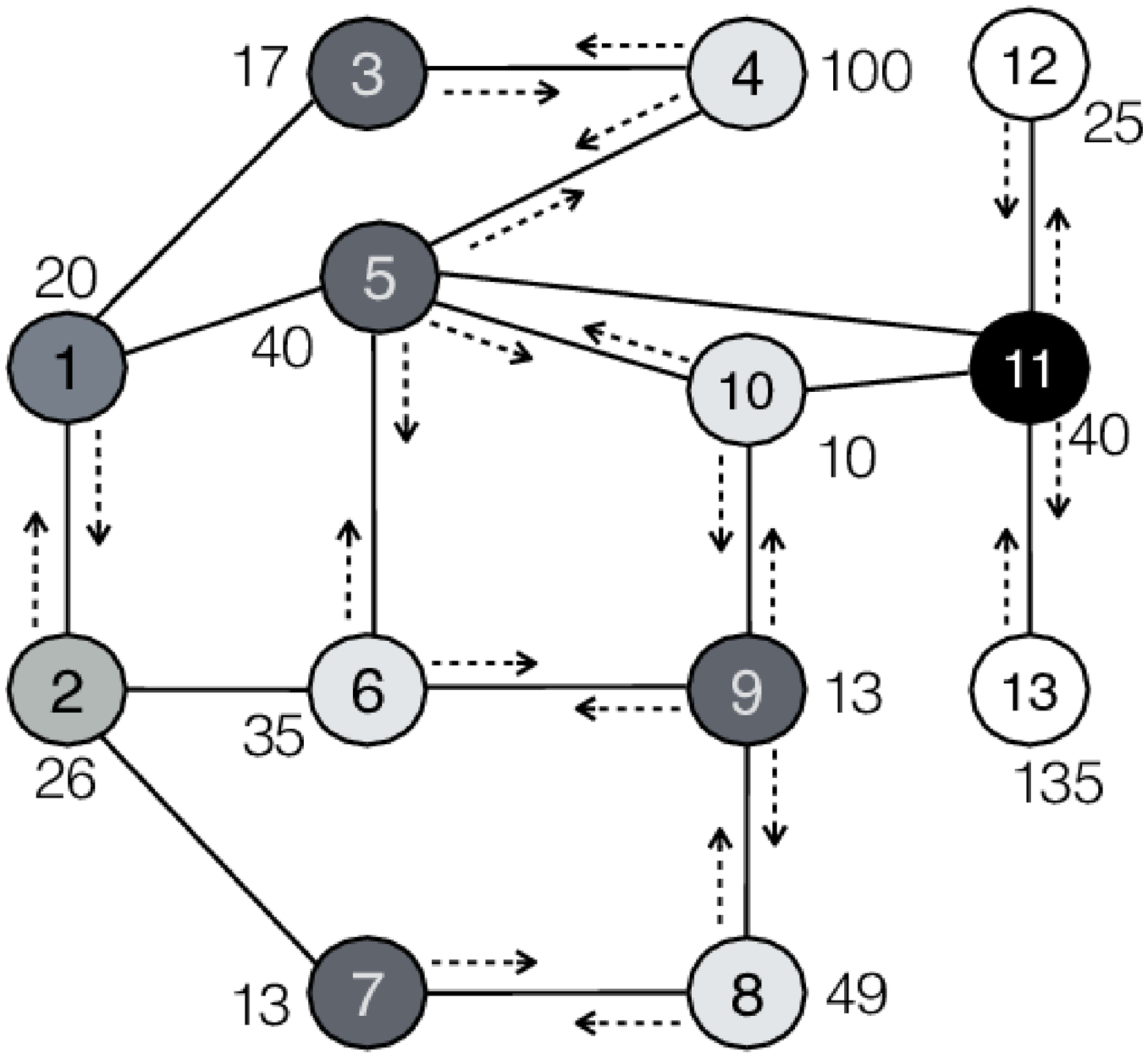}
\caption{}
\label{fig:simple-examples-2}
\end{subfigure}
\begin{subfigure}[b]{0.23\textwidth}	
\includegraphics[scale=0.2]{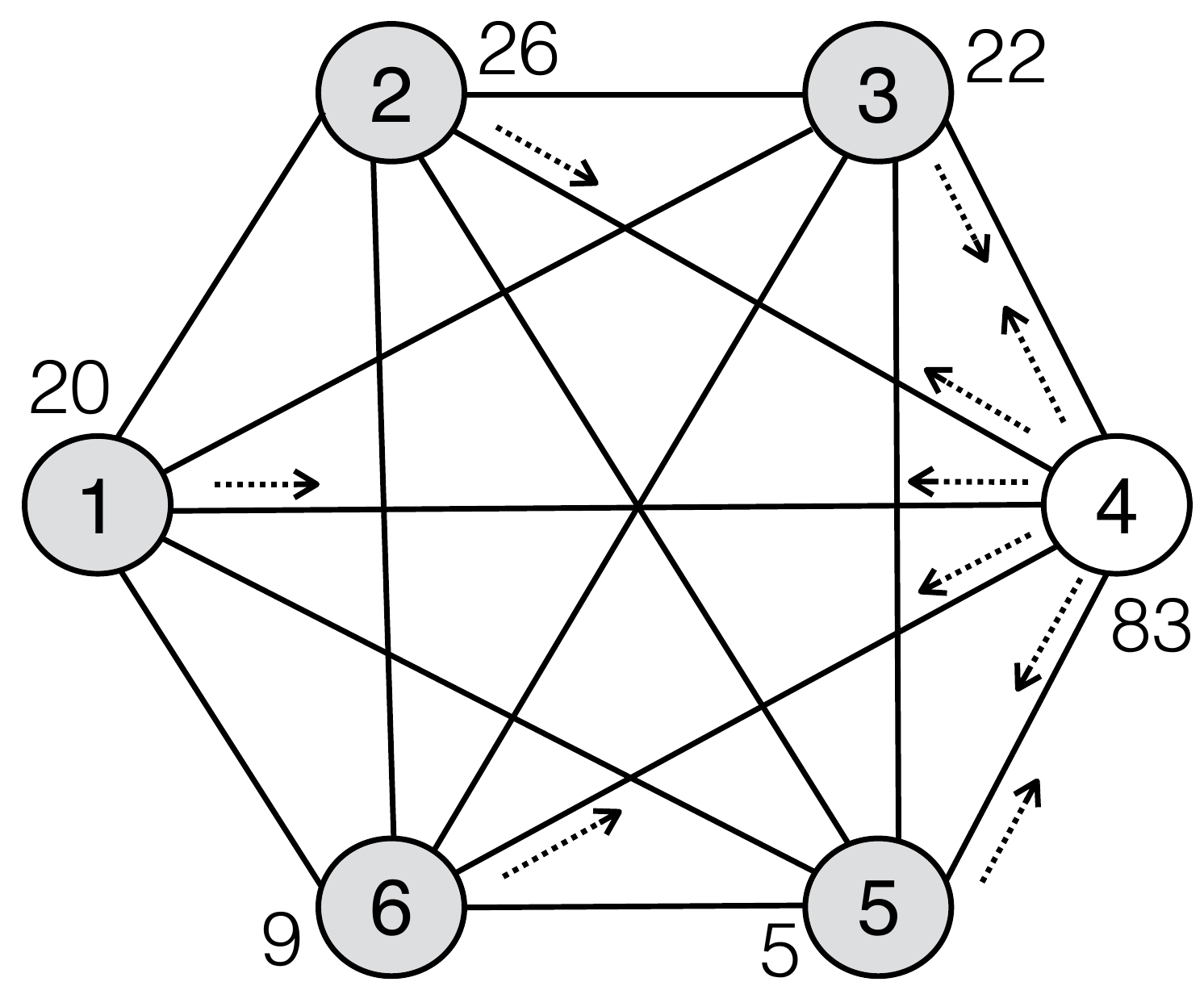}
\caption{}
\label{fig:simple-examples-3}	
\end{subfigure}
\begin{subfigure}[b]{0.23\textwidth}	
\includegraphics[scale=0.2]{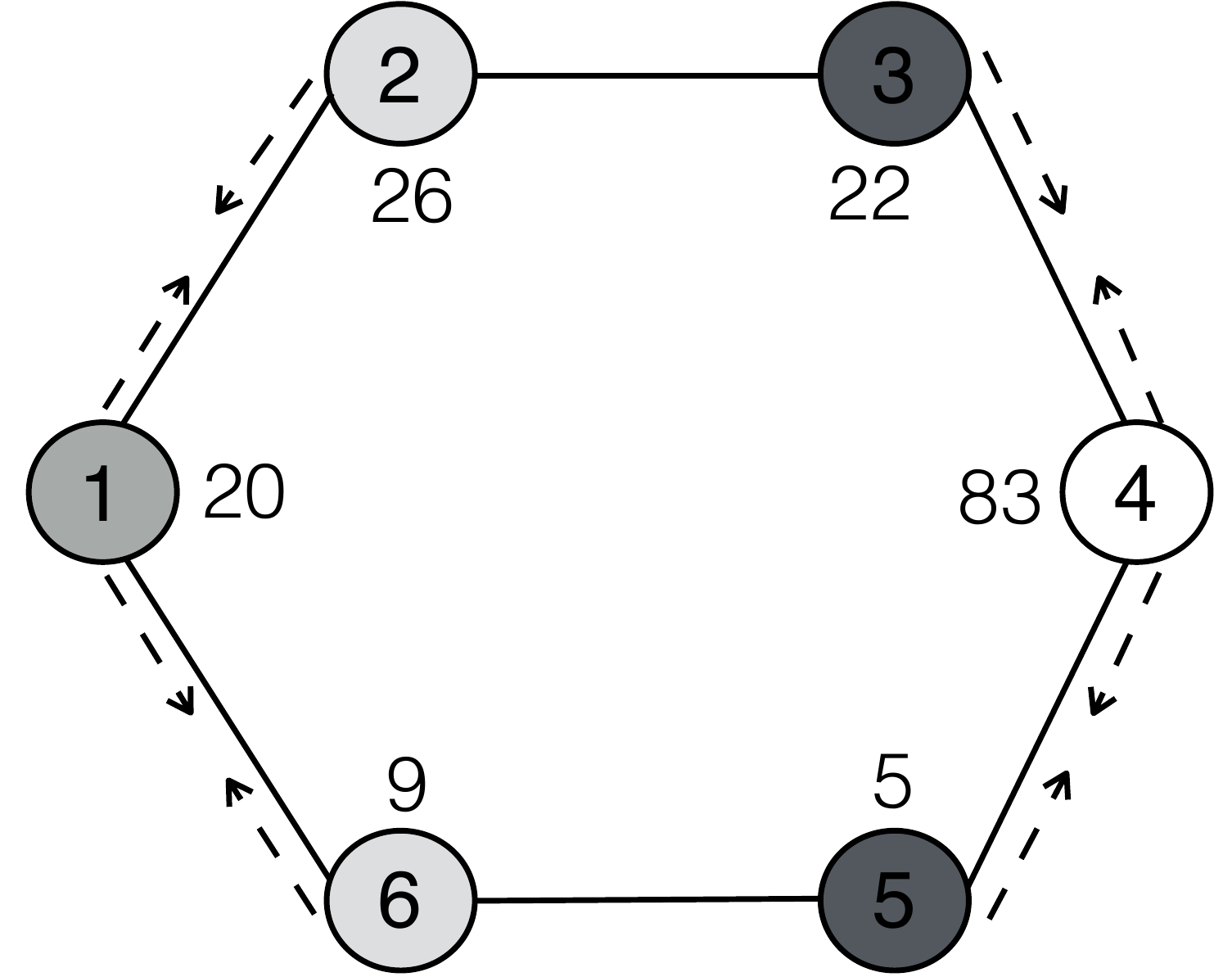}
\caption{}
\label{fig:simple-examples-4}
\end{subfigure}
\caption{\textbf{(a)}: A network with 6 nodes that create 2 groups, each one marked with the dotted-line rectangle. There are 3 different levels of sharing ratios. The color of each node is analogous to its sharing ratio value (increasing from white to black colour). The received resources are $r_{1}^{*}=20,$ $r_{2}^{*}=40,$ $r_{3}^{*}=10,$ $r_{4}^{*}=10$, $r_{5}^{*}=60,$ and $r_{6}^{*}=30$. The sharing ratios for the nodes belonging to each set are $u_{1}^{*}=0.5$, $u_{2}^{*}=1$, and $u_{3}^{*}=2$ respectively; \textbf{(b)}: A network with $13$ nodes which create $3$ groups. Received resources are $r_{1}^{*}=26$, $r_{2}^{*}=20$, $r_{3}^{*}=39.74$, $r_{4}^{*}=42.78$, $r_{5}^{*}=93.49$, $r_{6}^{*}=14.97$, $r_{7}^{*}=30.38$, $r_{8}^{*}=20.96$, $r_{9}^{*}=30.38$, $r_{10}^{*}=4.28$, $r_{11}^{*}=160$, $r_{12}^{*}=6.25$, and $r_{13}^{*}=33.75$. \textbf{(c)}: A ring graph of $6$ nodes with $2$ coalition and $4$ level. \label{fig:complete-graph-6nodesA-1} \textbf{(d)}: A complete graph of $6$ nodes with $1$ coalition and $2$ levels.}
\label{figure:simple-examples} 
\end{figure}

Consider first the networks of Fig. \ref{figure:simple-examples}. Solid lines represent the physical connections of each node, i.e., the possible exchanges that this sharing economy network can support, and the dotted arrows indicate the actual resource allocations that take place at the equilibrium point. Next to each node we depict its resource endowment. 

Let us first focus on the 6-node network of Fig. \ref{fig:simple-example-1}. At the lex-optimal equilibrium point this network has $K^{*}=3$ levels with 3 sets $\mathcal{L}_{1}^{*}={\{1,6\}},\,\mathcal{L}_{2}^{*}={\{3,4\}},\;\mathcal{L}_{3}^{*}={\{2,5\}}$ which are marked with different colors, where darker colors are used for nodes with higher sharing ratios. Let us now verify the properties that the lex-optimal allocation should have according to Theorem \ref{thm:MainTh0-1}. First, notice that set $\mathcal{L}_{1}^{*}$ is independent in graph $G$. Moreover, the neighbors of nodes in set $\mathcal{L}_{1}^{*}$ are the nodes in $\mathcal{L}_{3}^{*}$. Although nodes in $\mathcal{L}_{3}^{*}$ are physically connected, they only allocate resource to nodes in
$\mathcal{L}_{1}^{*}$ and it holds $\sum_{i\in\mathcal{L}_{3}^{*}}D_{i}=\sum_{i\in l_{1}}r_{i}=20+30$. Moreover, the highest and the lowest levels satisfy the condition $u_{1}^{*}u_{3}^{*}=1$. The nodes are partitioned into $2$ disjoint groups $\mathcal{M}_{1}^{*}=\mathcal{L}_{1}^{*}\cup\mathcal{L}_{3}^{*}$ and $\mathcal{M}_{2}^{*}=\mathcal{L}_{2}^{*}$, each one containing
nodes with at most two levels. 

For the example of Fig. \ref{fig:simple-examples-2} we used a network with $13$ nodes that yields $K^{*}=6$ levels, with $u_{1}^{*}=0.25$, $u_{2}^{*}=0.43$, $u_{3}^{*}=0.77$, $u_{4}^{*}=2.34$, $u_{5}^{*}=1.3$, and $u_{6}^{*}=4$. The sets are $\mathcal{L}_{1}^{*}=\{12,\,13\}$, $\mathcal{L}_{2}^{*}=\{4,\,6,\,8,\,10\}$, $\mathcal{L}_{3}^{*}=\{2\}$,
$\mathcal{L}_{4}^{*}=\{1\}$, $\mathcal{L}_{5}^{*}=\{3,\,5,\,7,\,9\}$, and $\mathcal{L}_{6}^{*}=\{11\}$. Sets $\mathcal{L}_{1}^{*}$, $\mathcal{L}_{2}^{*}$, and $\mathcal{L}_{3}^{*}$ are independent in graphs $G_{\mathcal{Q}_{1}},$ $G_{\mathcal{Q}_{2}}$, $G_{\mathcal{Q}_{3}}$, and the set $\mathcal{L}_{1}^{*}\cup\mathcal{L}_{2}^{*}\cup\mathcal{L}_{3}^{*}$ is independent in $G$. Moreover, it holds $\mathcal{L}_{6}^{*}=\mathcal{N}_{\mathcal{Q}_{1}}(\mathcal{L}_{1}^{*})$, $\mathcal{L}_{5}^{*}=\mathcal{N}_{\mathcal{Q}_{2}}(\mathcal{L}_{2}^{*})$ and $\mathcal{L}_{4}^{*}=\mathcal{N}_{\mathcal{Q}_{3}}(\mathcal{L}_{3}^{*})$, and holds $u_{6}^{*}u_{1}^{*}=u_{5}^{*}u_{2}^{*}=u_{4}^{*}u_{3}^{*}=1$. In this example we have 3 disjoint groups $\mathcal{M}_{1}^{*}=\mathcal{L}_{1}^{*}\cup\mathcal{L}_{6}^{*}$,
$\mathcal{M}_{2}^{*}=\mathcal{L}_{2}^{*}\cup\mathcal{L}_{5}^{*}$, and $\mathcal{M}_{3}^{*}=\mathcal{L}_{3}^{*}\cup\mathcal{L}_{4}^{*}$. We see that links $(10,11)$, $(5,11)$, $(1,3)$, $(1,5)$ and $(2,7)$
are redundant and can be removed without affecting the lex-optimal allocation.


\begin{figure}
\centering
\begin{subfigure}[b]{0.3\textwidth}
\includegraphics[scale=0.35]{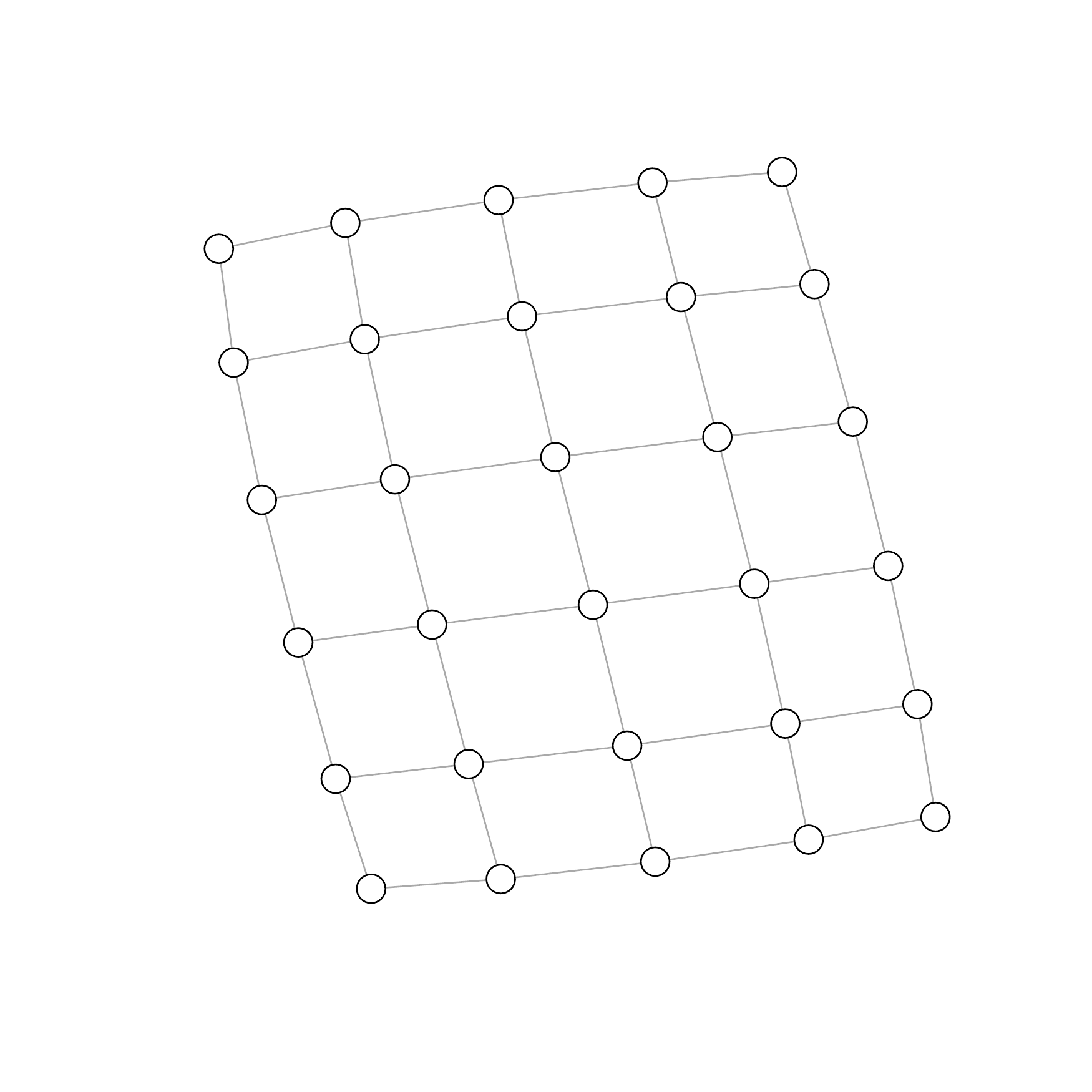}
\caption{}
\label{fig:lattice-homogeneous}
\end{subfigure}
\begin{subfigure}[b]{0.3\textwidth}
\includegraphics[scale=0.35]{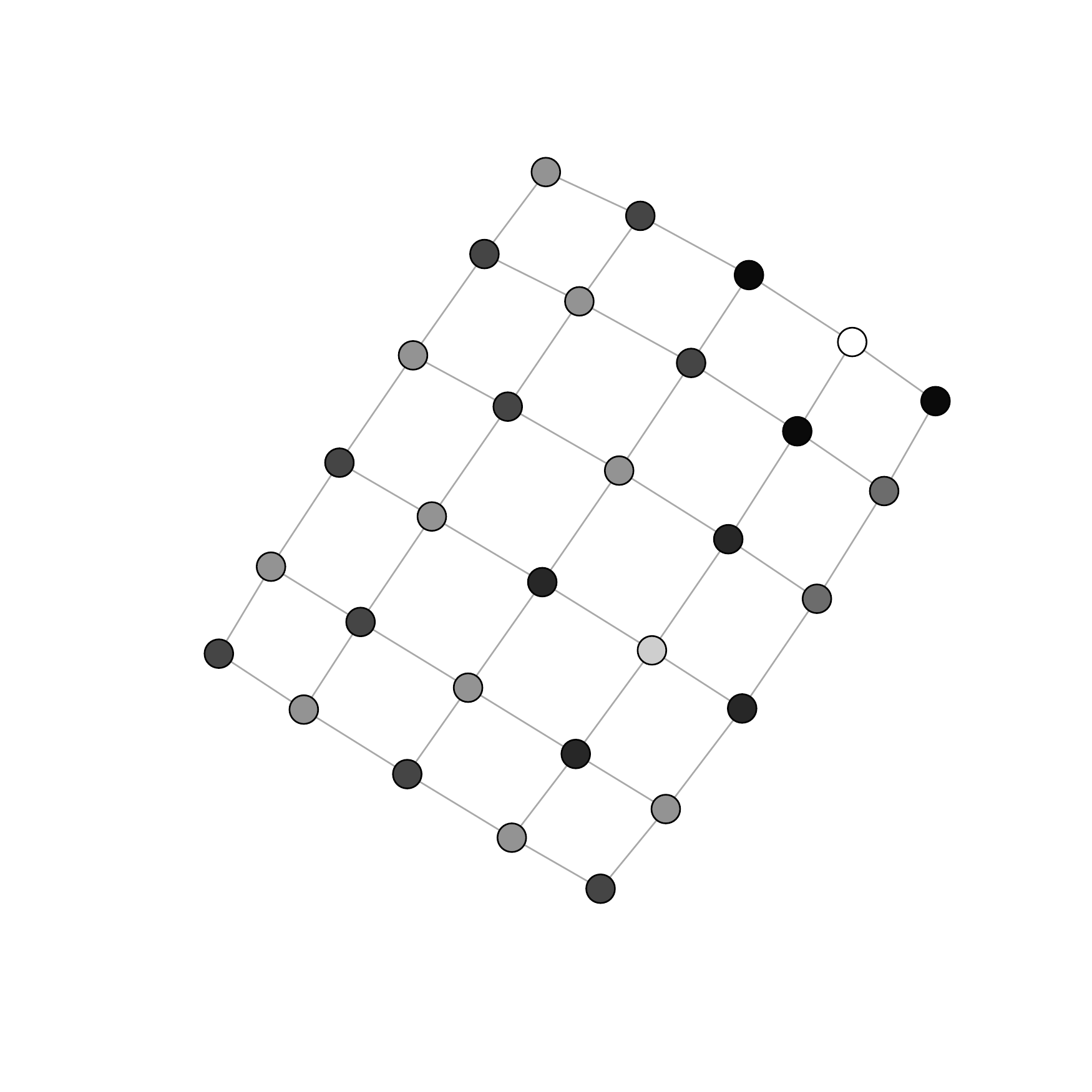}		
\caption{}
\label{fig:lattice-diverse-with-2}	
\end{subfigure}
\begin{subfigure}[b]{0.3\textwidth}
\includegraphics[scale=0.35]{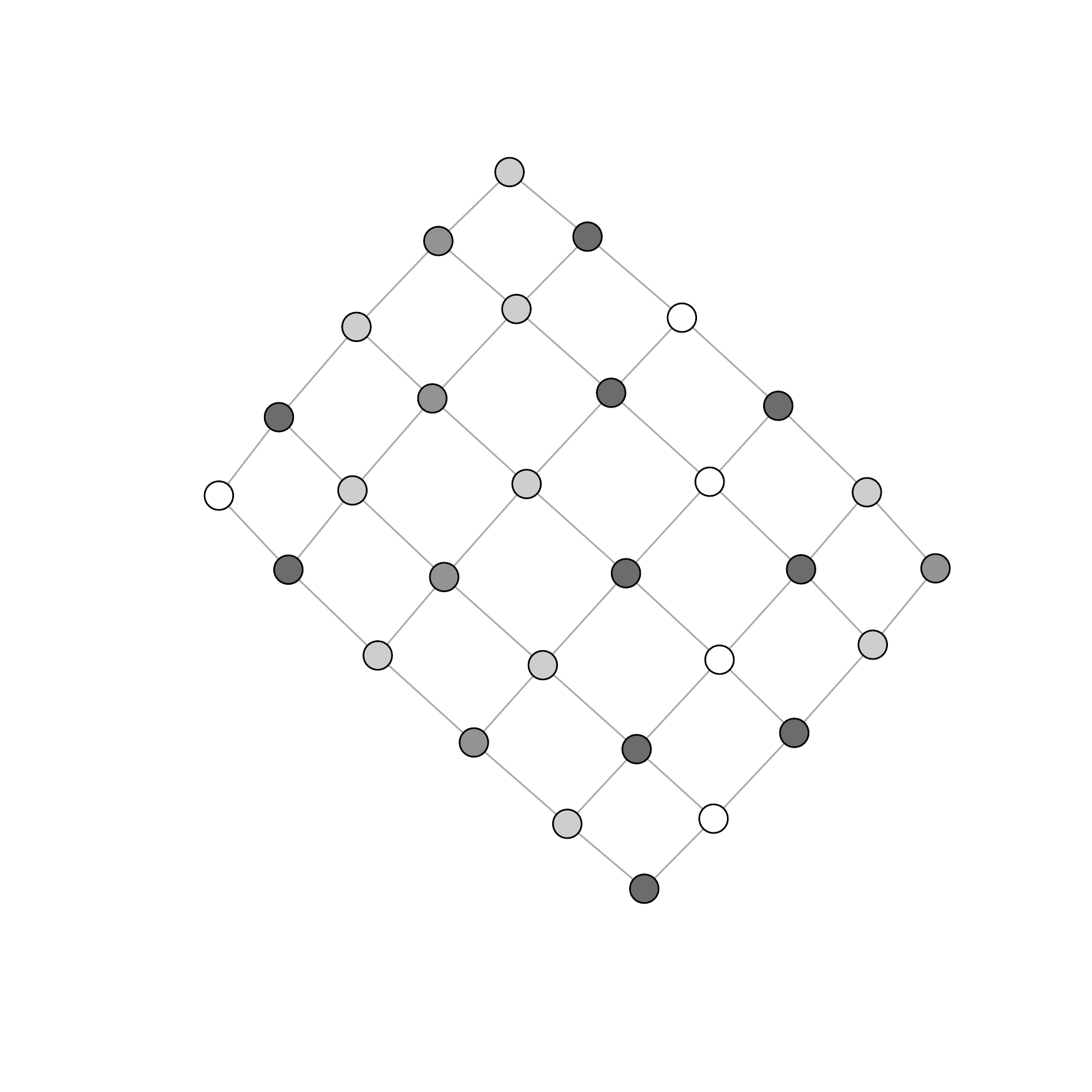}
\caption{}
\label{fig:lattices-diverse-with-5}
\end{subfigure}	
\caption{Lattice Graphs with N=30 nodes. (a): $N=30$ equal-endowment nodes ($D=30$), 1 sharing ratio level. (b): $28$ low-endoment nodes ($D=30$), 2 high-endowment nodes ($D=300$), 7 different sharing ratio levels. (c): $25$ low-endoment nodes (30), 5 high-endowment nodes ($D=300$), 4 different	sharing ratio levels.}
\label{fig:lattices-main}
\end{figure}

\begin{figure}
\centering
\begin{subfigure}[b]{0.3\textwidth}
\includegraphics[scale=0.4]{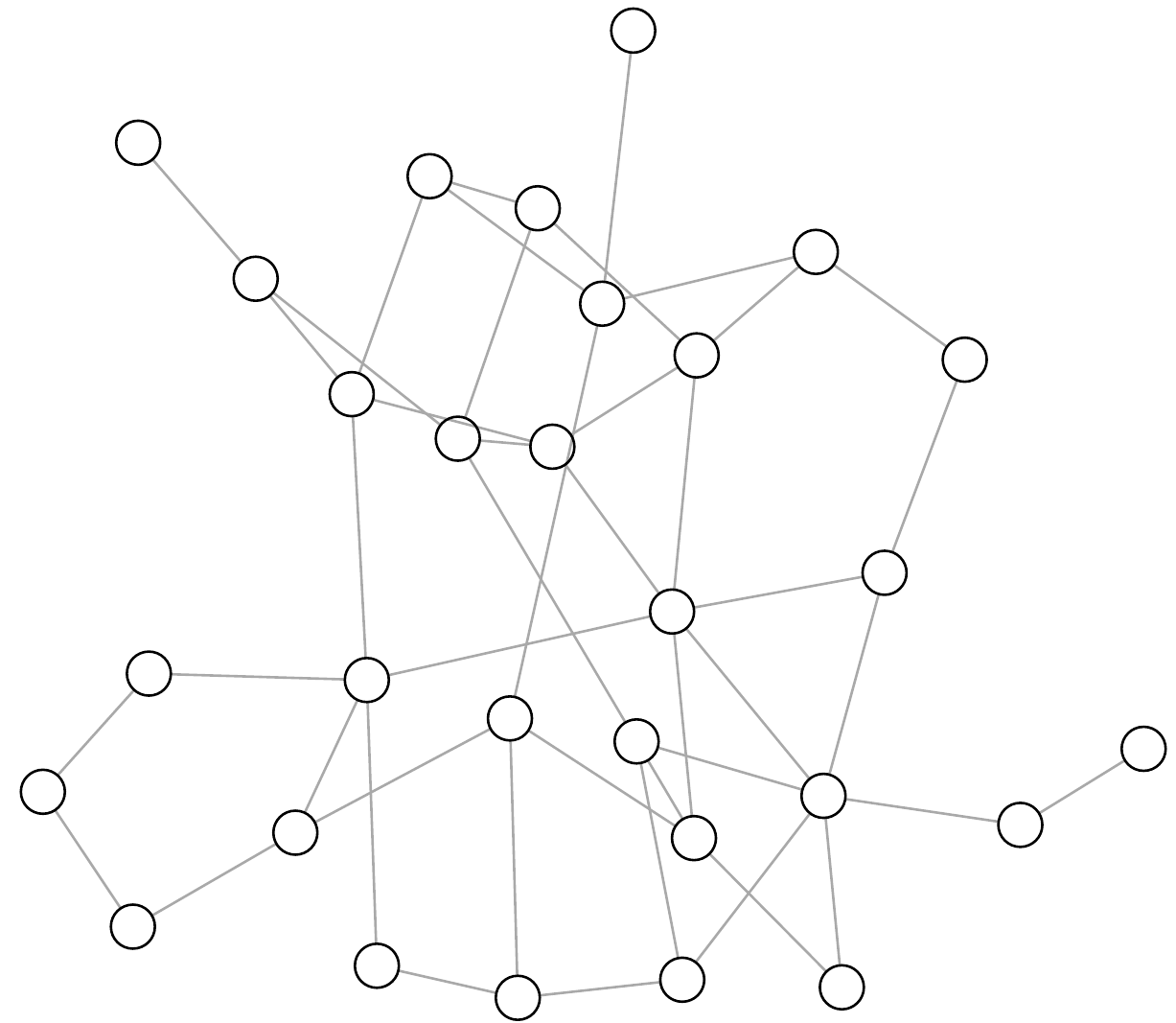}	
\caption{}
\label{fig:E-R-homogeneous-01}
\end{subfigure}	
\begin{subfigure}[b]{0.3\textwidth}
\includegraphics[scale=0.4]{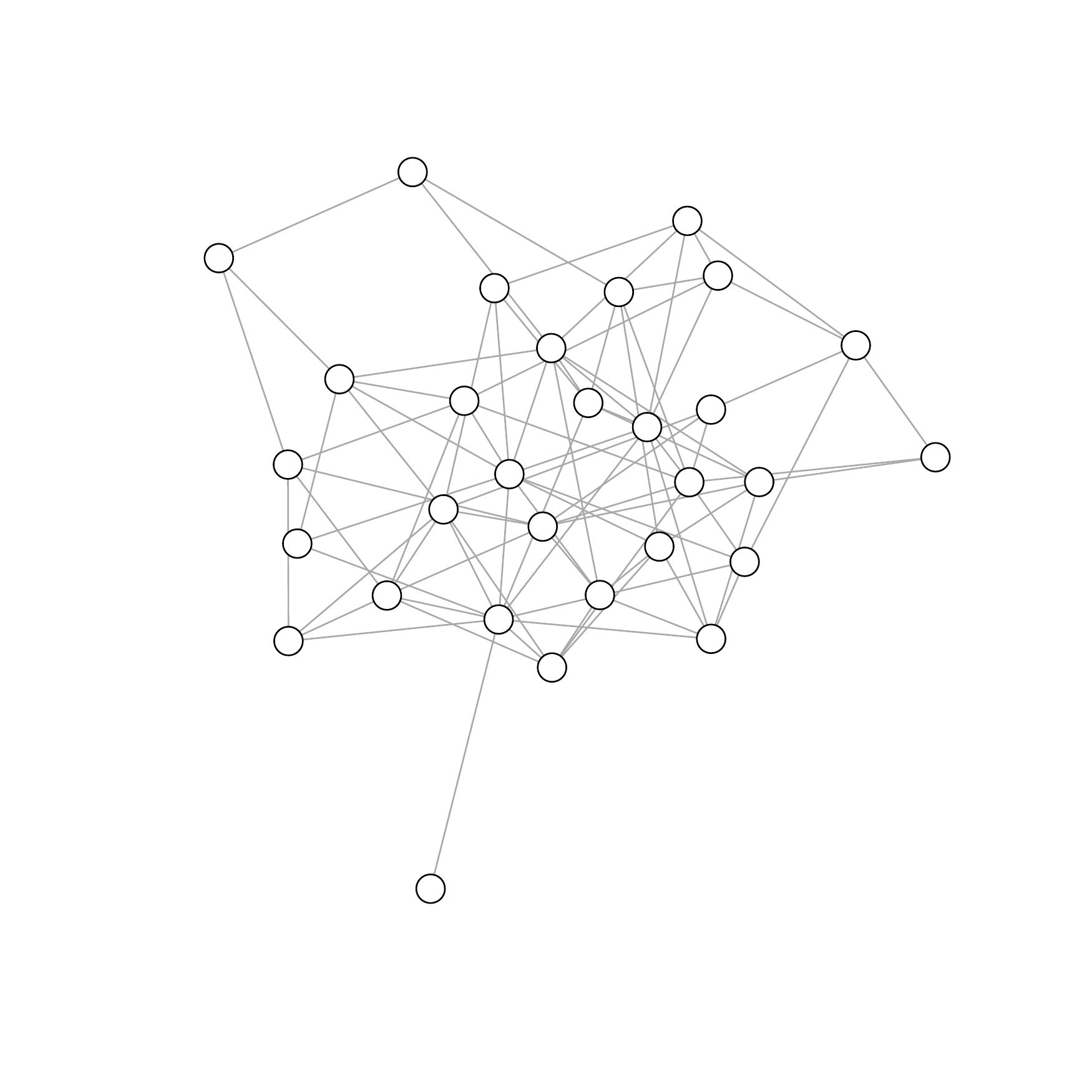}
\caption{}
\label{fig:ER-homogeneous-02}
\end{subfigure}	
\begin{subfigure}[b]{0.3\textwidth}	
\includegraphics[scale=0.4]{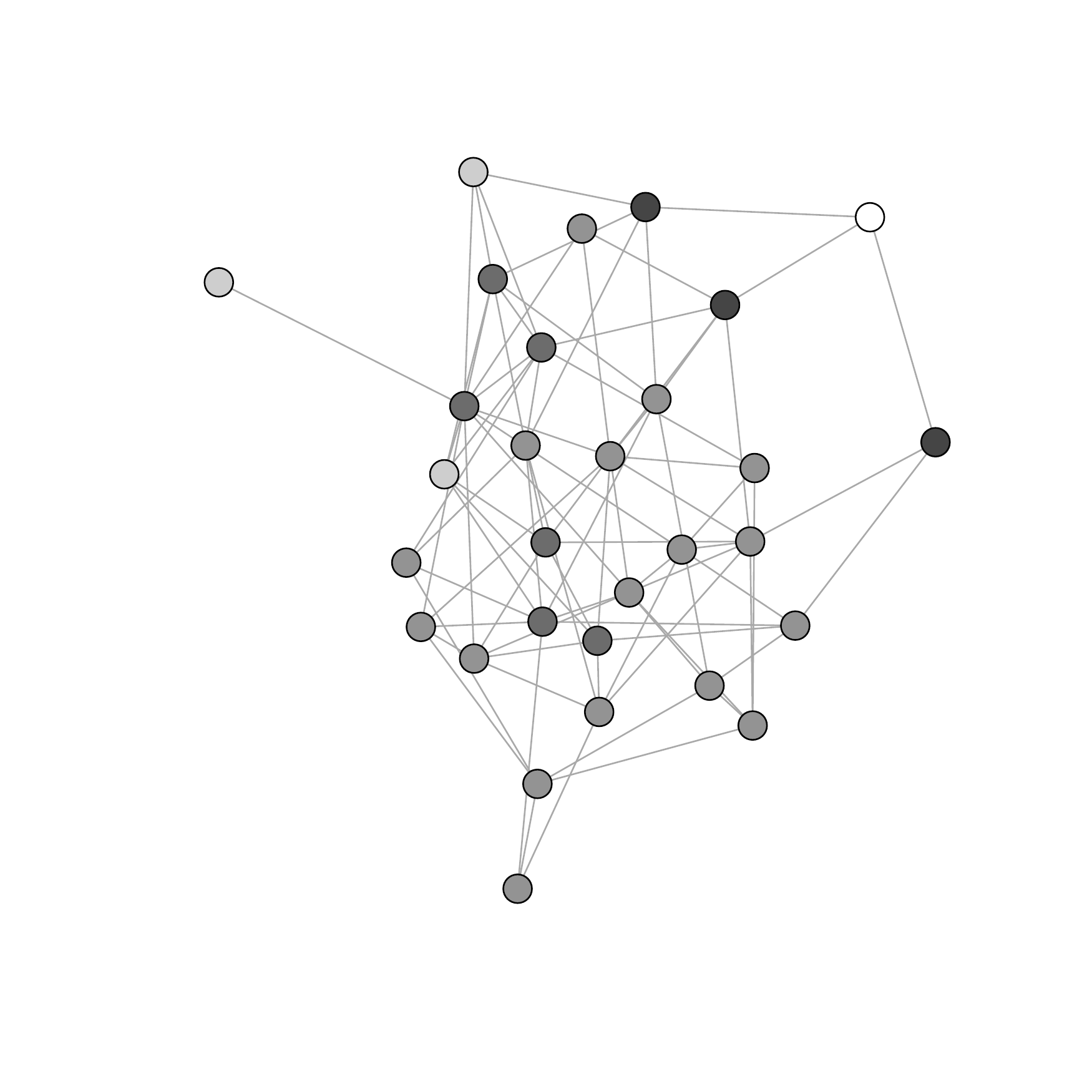}	
\caption{}
\label{fig:ER-diverse-02}
\end{subfigure}	
\caption{Equilibriums in Erdos-Renyi Graphs. (a): $N=30$, link creation probability $p=0.1$, 1 sharing ratio level. (b): $N=30$, $p=0.2$, 1 sharing ratio level. (c): $p=0.2$, $2$ high-endowment nodes ($D=300$), 4 sharing ratio levels.}
\end{figure}

\begin{figure}
\centering
\begin{subfigure}[b]{0.3\textwidth}	
\includegraphics[scale=0.4]{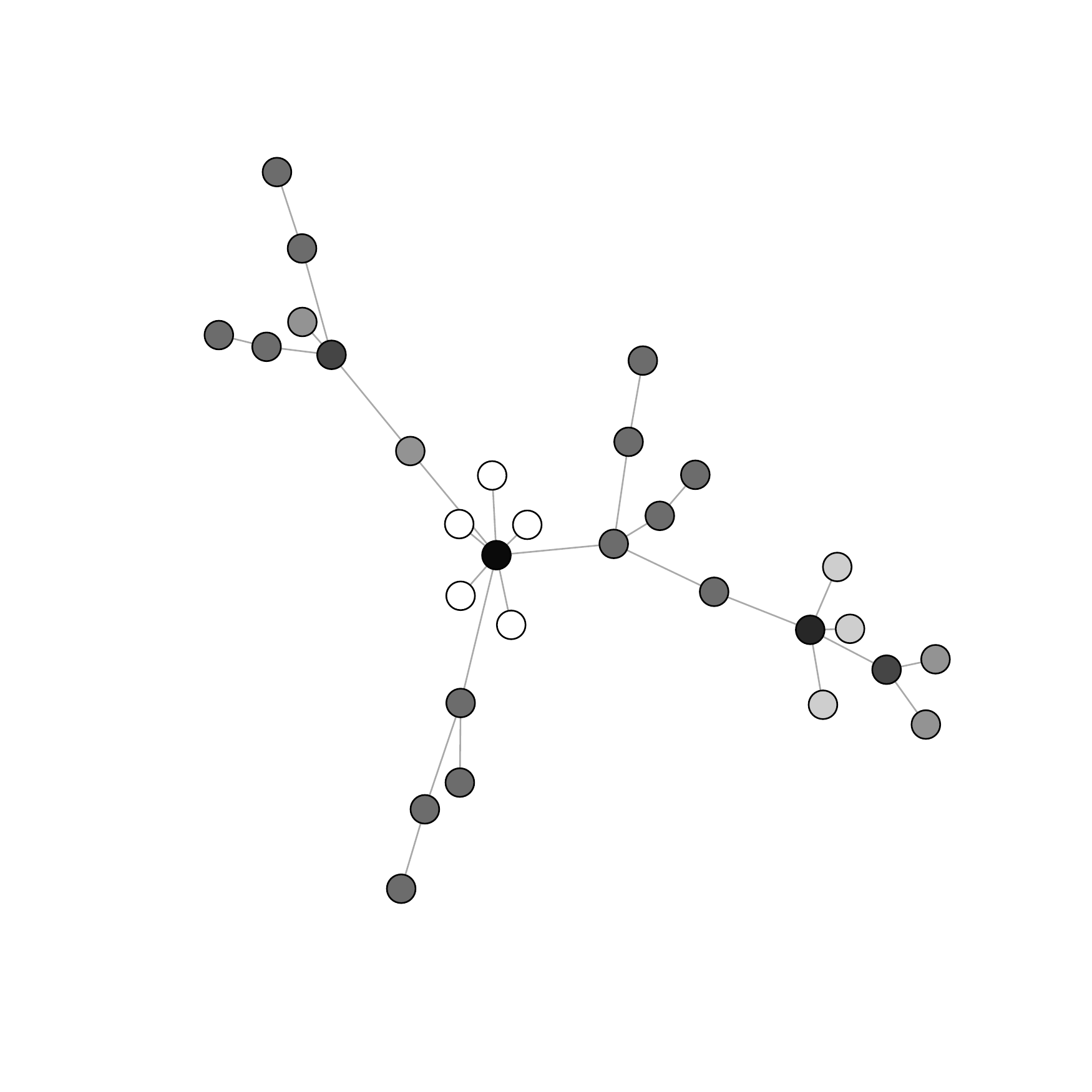}
\caption{}
\label{fig:scale-free-05}
\end{subfigure}	
\begin{subfigure}[b]{0.3\textwidth}		
\includegraphics[scale=0.4]{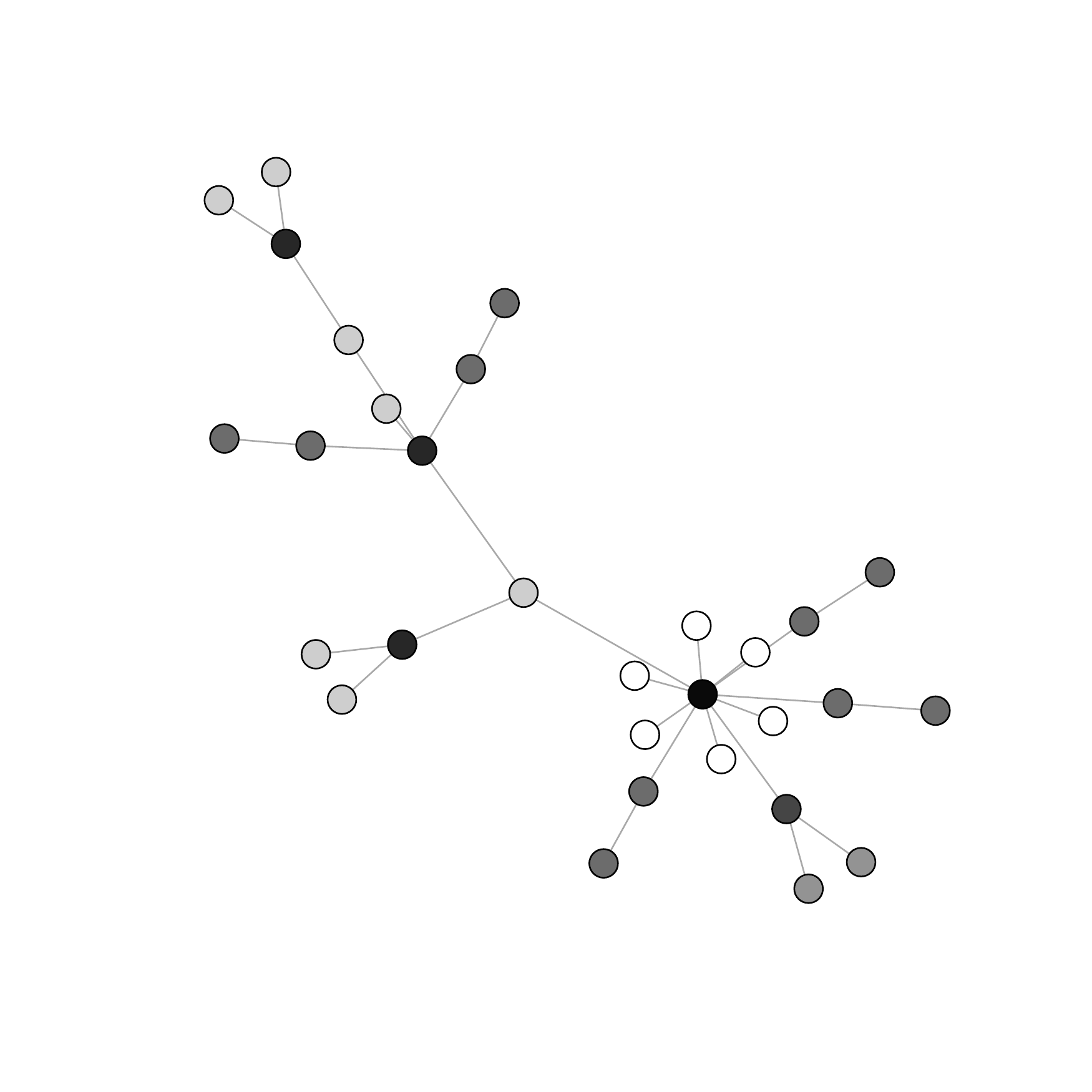}
\caption{}
\label{fig:scale-free-1}
\end{subfigure}	
\begin{subfigure}[b]{0.3\textwidth}			
\includegraphics[scale=0.4]{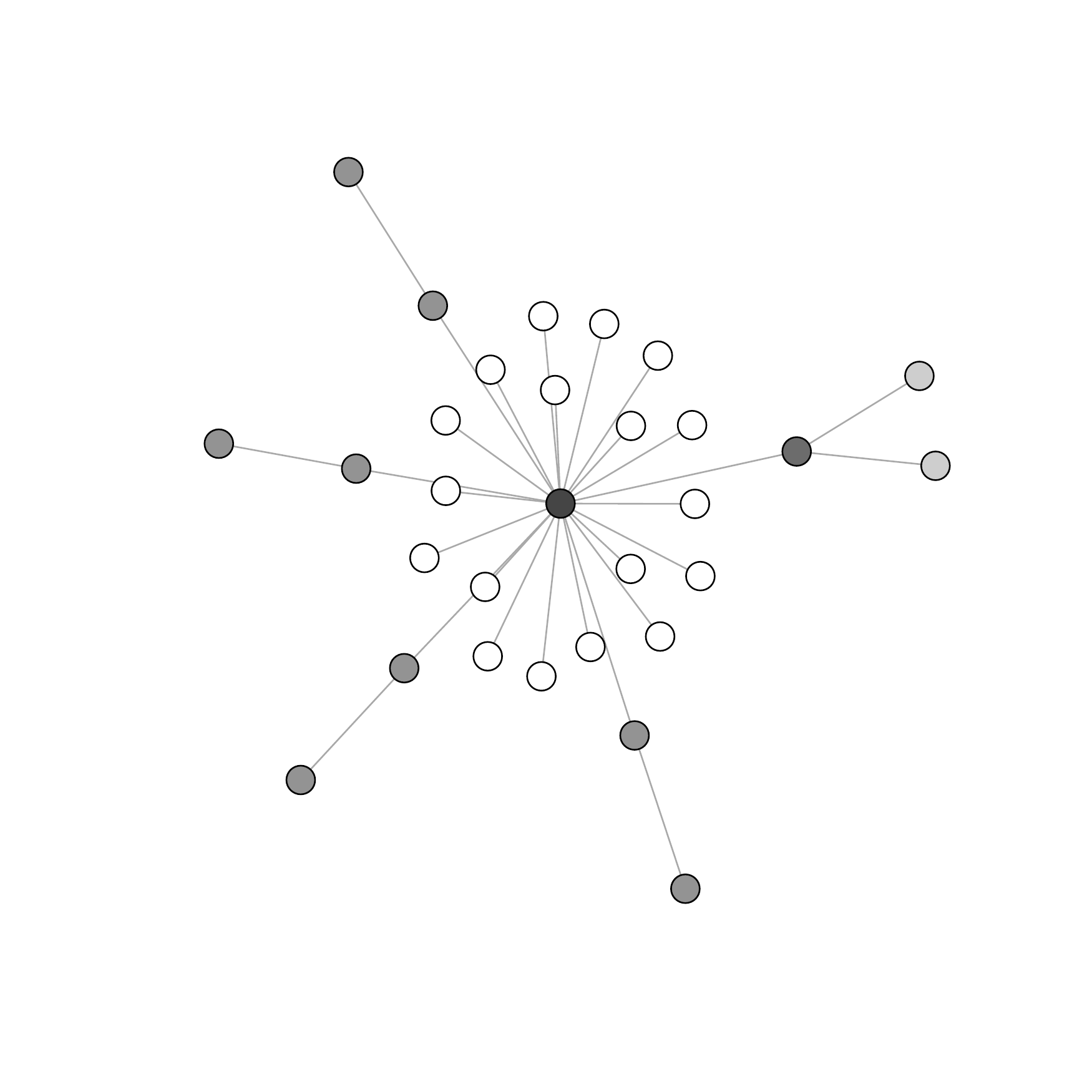}
\caption{}
\label{fig:scale-free-2}
\end{subfigure}	
\caption{Scale-free Graphs. (a): low skewed with skew parameter 0.5, 5 sharing ratio levels. (b): moderately skewed (linear model, skew parameter 1), 7 sharing ratio levels. (c): highly skewed (parameter 2), 5 sharing ratio levels. }
\end{figure}

\begin{figure}
\centering
\begin{subfigure}[b]{0.3\textwidth}			
\includegraphics[scale=0.4]{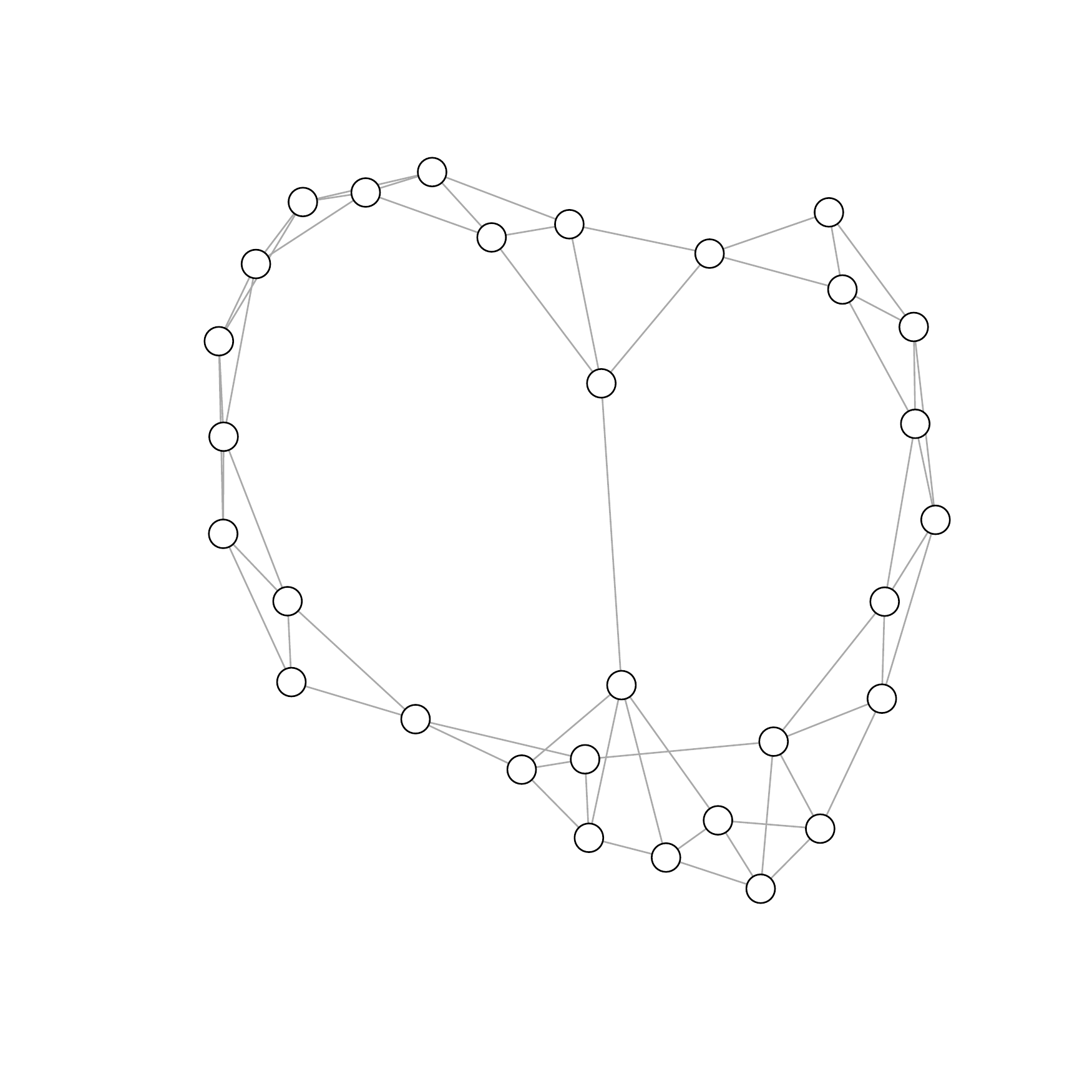}
\caption{}
\label{fig:small-world-symmetric}
\end{subfigure}			\qquad\qquad
\begin{subfigure}[b]{0.3\textwidth}	
\includegraphics[scale=0.4]{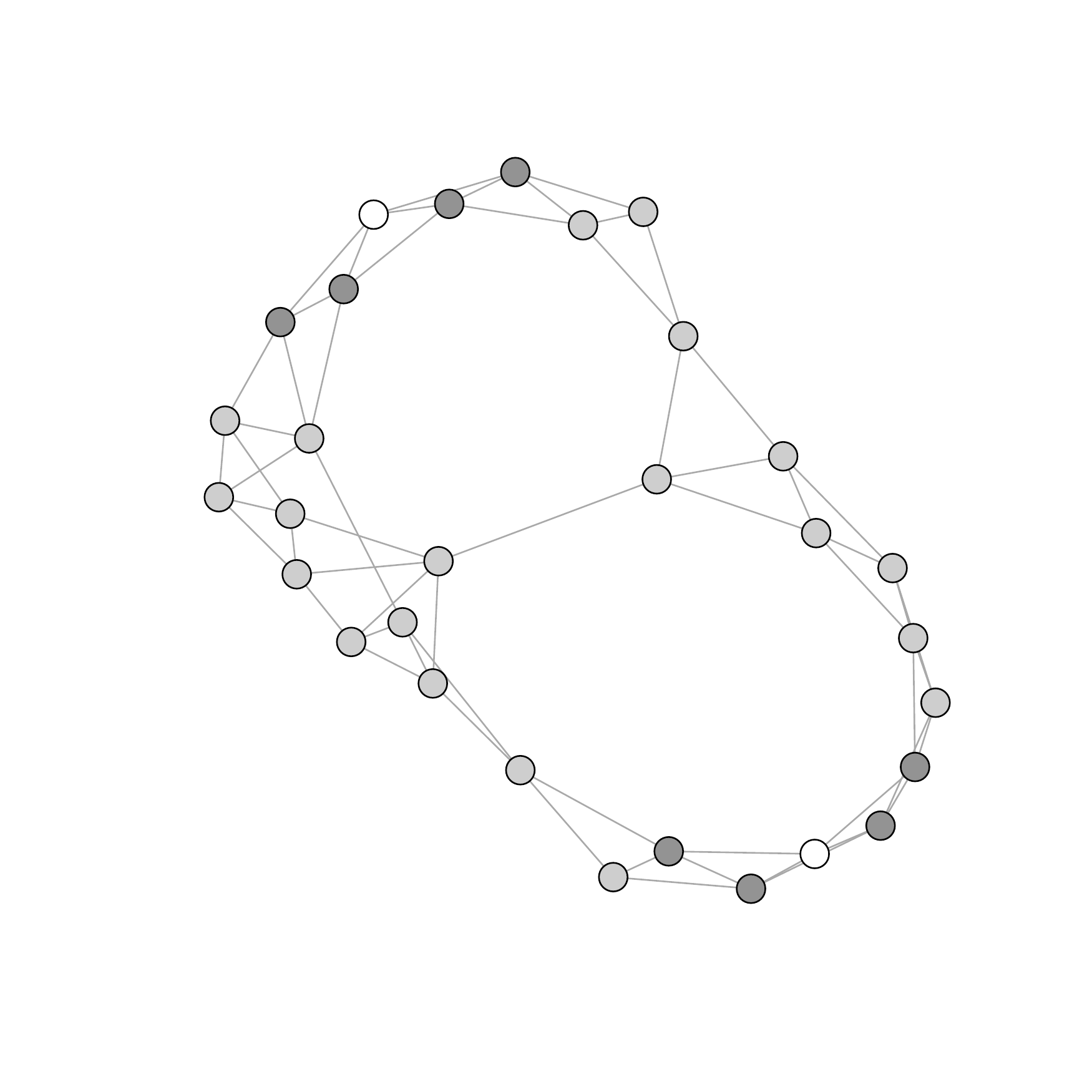}
\caption{}		
\label{fig:small-world-asymmetric}
\end{subfigure}	
\caption{Small-world Graph, N=30. (a):Small-world Graph with symmetric nodes; (b):Small-world Graph with Asymmetric Endowments (4 nodes with D=300)}	
\label{fig:small-world-main-figure}
\end{figure}

Figure \ref{fig:simple-examples-3} depicts a complete graph with $6$ nodes, where node $i=4$ has level $u_{I_{1}(r^{*})}=0.988$
while the other nodes have level $u_{I_{2}(r^{*})}=1.012$. In general for complete graphs, from Theorem \ref{thm:MainTh0-1} and the fact that independent sets in such graphs contain only one node, it follows that lex-optimal allocations may have at most two levels. Moreover a complete graph has two levels iff the resource of node $i_{0}$ with the maximum endowment is larger than the sum of the resources of the rest of the nodes, and it is $\mathcal{L}_{1}=\{i_{0}\}$. On the other hand, for the respective $6$-node ring graph, Fig. \ref{fig:simple-examples-4}, the lex-optimal solution yields $4$ levels. 

\subsection{Typical Network Models}

In this subsection we focus on larger graphs of typical models, namely the Lattice, Erdos-Renyi \citep{erdos-renyi}, Scale-free \citep{scale-free}, and Small-world \citep{small-world} networks. We demonstrate that the reached equilibrium points can be affected by the properties of these networks, e.g., their density, but also by the diversity in the nodes' resource endowments. In Figure \ref{fig:lattices-main} we present the equilibrium allocations for 3 lattice graphs with 30 nodes. First, Figure \ref{fig:lattice-homogeneous} presents the homogeneous case where every node has an average generation rate of $D_{i}=30$ resource units. The node color is modulated by the attained equilibrium sharing ratio, with darker shades indicated higher sharing ratios. We observe first that in the homogeneous case, the graph structure does not create any imbalance in the equilibrium, all nodes achieve an sharing ratio equal to 1. This result changes significantly when the nodes have diverse resource endowments. Namely, in Figure \ref{fig:lattice-diverse-with-2} we depict the equilibrium of the same lattice network where 2 out of the 30 nodes have now much higher resource, namely $D_{i}=300$. This creates 7 different sharing ratio levels. Finally, Figure \ref{fig:lattices-diverse-with-5}depicts a graph where 5 out of the 30 nodes have resource $D_{i}=300$, and this creates 4 different sharing ratio levels, making the sharing economy network less imbalanced in that respect.

Next we focus on Erdos-Renyi (E-R) graphs. Figure \ref{fig:E-R-homogeneous-01} depicts an E-R graph with $N=30$ nodes and link creation probability $p=0.1$, where all nodes are homogeneous in terms of their resource endowments, having $D_{i}=30$. We observe that all nodes reach the same unique equilibrium point of equal exchanges. The same holds for Figure \ref{fig:ER-homogeneous-02} which is denser ($p=0.2$), yet the additional links do not affect the allocation strategies of the nodes. However, when the nodes become resource-diverse with 2 nodes having $D_{i}=300$, the equilibrium changes significantly. Namely, there are 5 sharing ratio levels with the minimum of them being as low as $u_{1}=0.33$ and the maximum as high as $u_{5}=3.33$. In other words, similarly to the lattice graph we observe that a change in the resource endowments is more likely to change the equilibrium than a change in the structure of these networks.

This is not the case however for scale-free graphs. These networks which are formed through preferential-attachment processes, do not show assortative mixing \citep{newman-assortative}, and specifically many nodes with low degree are connected to nodes with much higher degree. This creates a structural advantage which results in diverse equilibrium sharing ratios even for homogeneous systems where nodes have identical resource endowments. Figure \ref{fig:scale-free-05} presents the equilibrium in a scale-free graph that has a power parameter $k=0.5$. We observe that there are 5 different sharing ratio levels, as some nodes have high degree while many others have degree 1. Figure \ref{fig:scale-free-1} presents a graph created by a linear preferential attachment process, i.e., $k=1$, that attains an equilibrium with 7 sharing ratio levels. Finally, for the graph of Figure \ref{fig:scale-free-2} it is $k=2$ and the equilibrium again changes and has 5 sharing ratio levels. In summary, we see with this basic example that in scale-free
graphs the equilibrium is significantly affected by the structural properties of the network graph and result in asymmetric points even when the nodes are identical in terms of their resource endowments.

As a final example, we present in Figure \ref{fig:small-world-main-figure} the equilibriums in 30-nodes graphs that has the small-world property \citep{small-world}. Figure \ref{fig:scale-free-1} presents a homogeneous network where all nodes have equal resource endowments (30 units) while Figure \ref{fig:scale-free-2} depicts the same network where 4 nodes have 300 units of average resource. This creates 3 different sharing ratio levels, instead of a single sharing ratio level for the former case. 

\begin{figure}
\centering
\begin{subfigure}[b]{0.32\textwidth}
\includegraphics[scale=0.25]{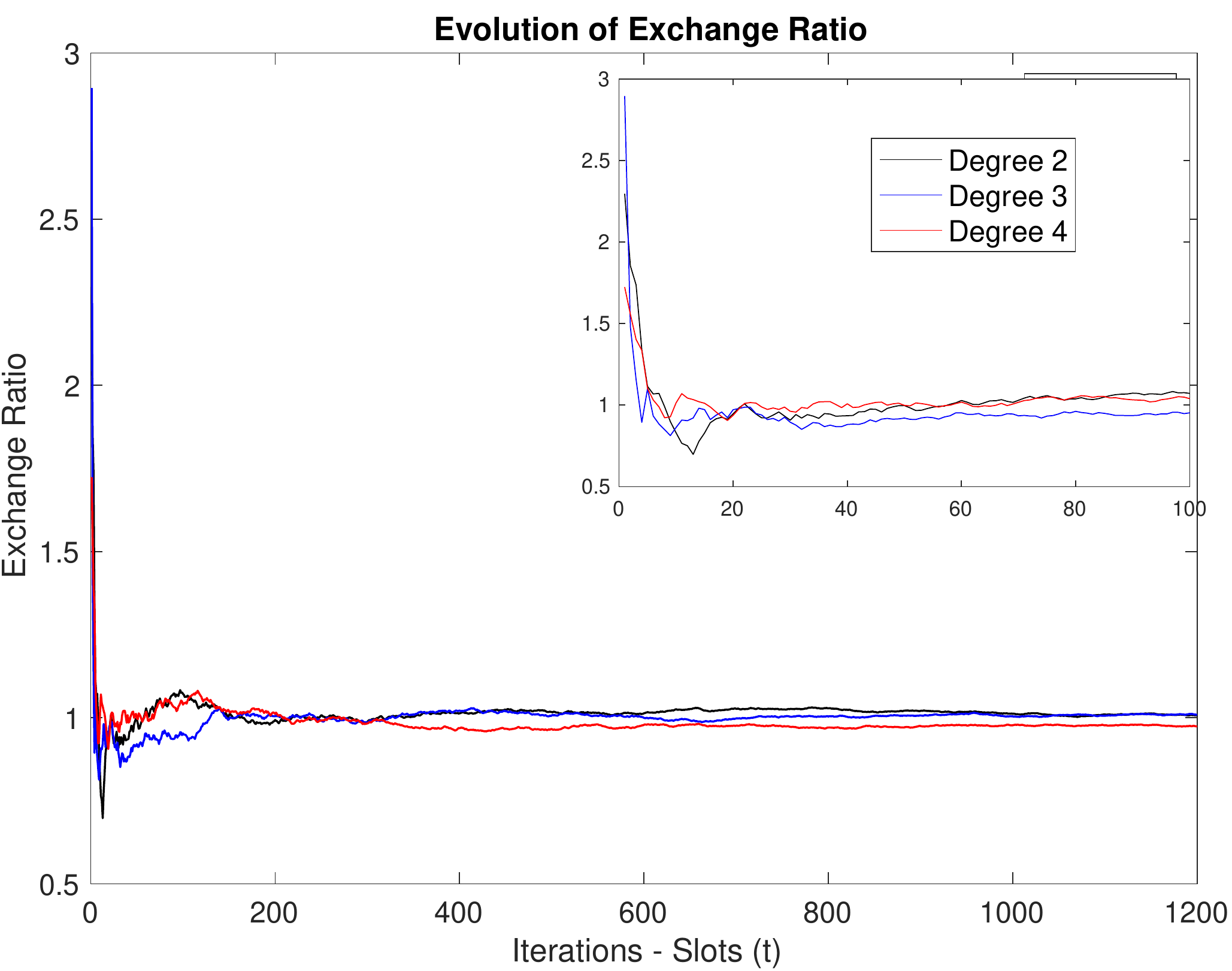}	
\caption{}
\label{fig:convergence-lattice}
\end{subfigure}	
\begin{subfigure}[b]{0.32\textwidth}
\includegraphics[scale=0.25]{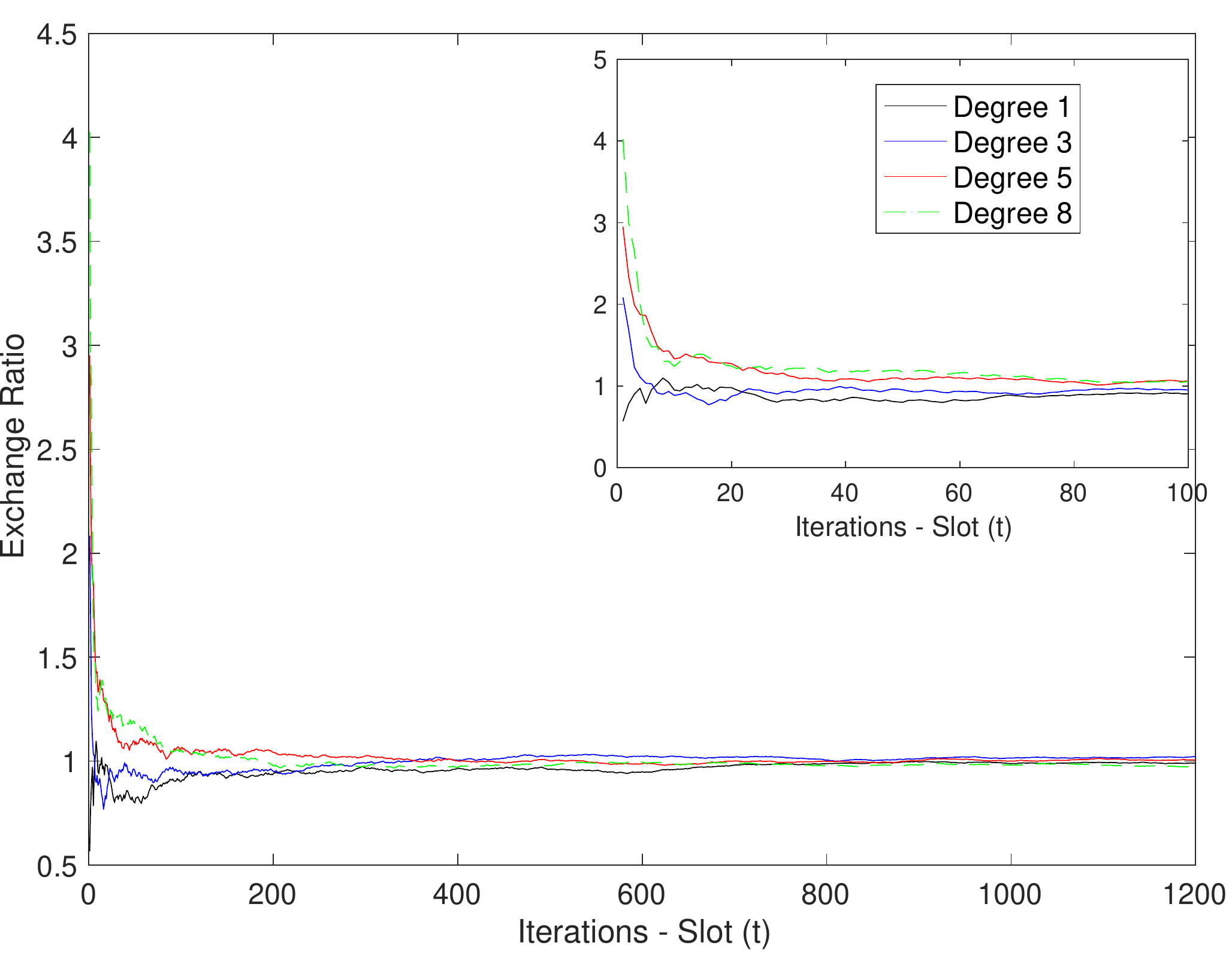}
\caption{}
\label{fig:convergence-erdos-renyi}
\end{subfigure}	
\begin{subfigure}[b]{0.32\textwidth}
{\includegraphics[scale=0.25]{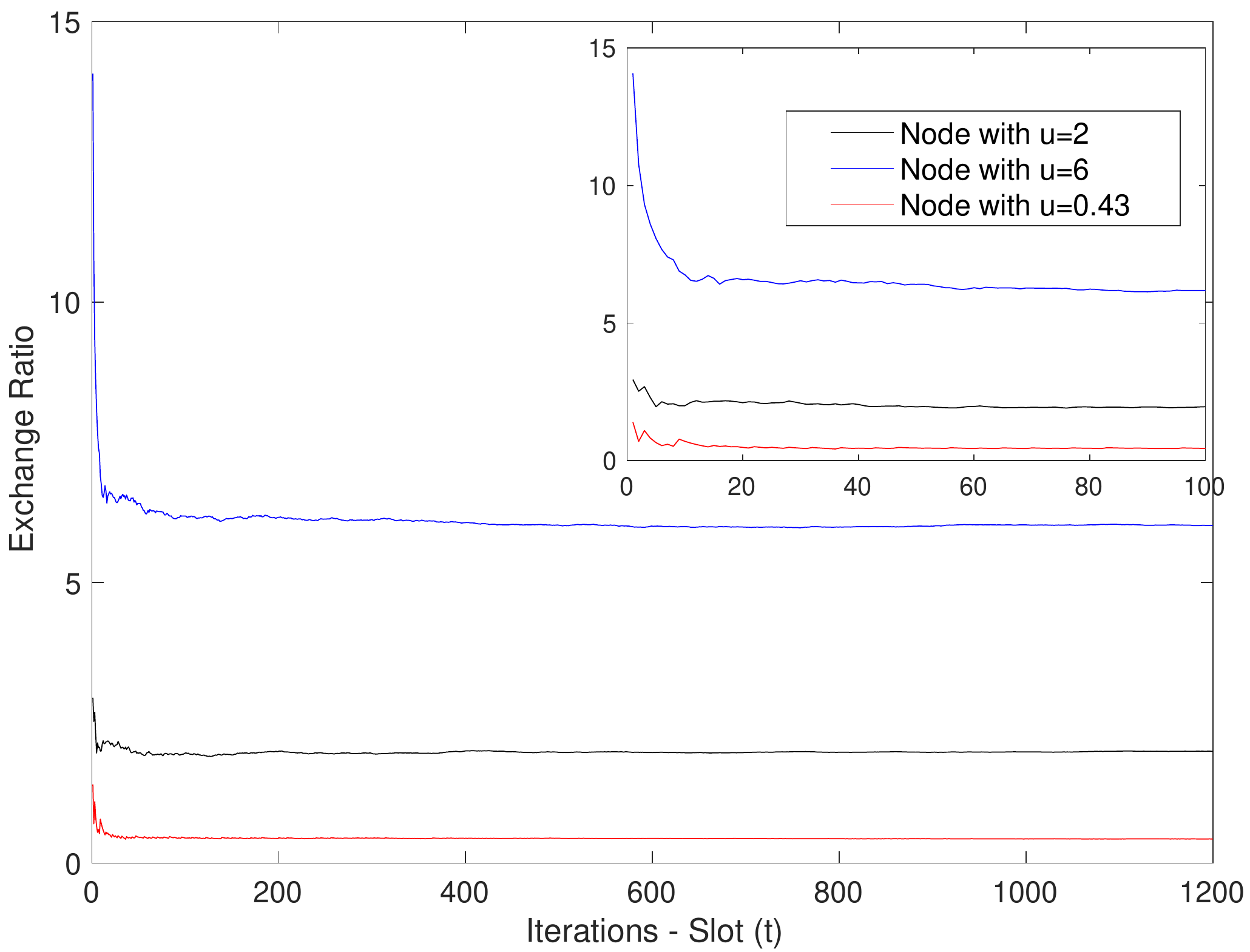}		
\caption{}
\label{fig:convergence-barabasi}}	
\end{subfigure}	
\label{fig:convergence-general}
\caption{Convergence of Algorithm 1. (a):Convergence of sharing ratios in the Lattice Network of Fig. \ref{fig:lattice-homogeneous}; results shown for 3 nodes with degrees 2, 3, and 4. Inset plot shows the first 100 iterations. (b): Convergence of sharing ratios in the Erdos-Renyi Network of Fig. \ref{fig:E-R-homogeneous-01}; results shown for 4 nodes with degrees 1, 3, 5 and 8. Inset plot shows the first 100 iterations. (c): Convergence of sharing ratios in the Scale-free Network of Fig. \ref{fig:scale-free-1}; results shown for 3 nodes with 3 different sharing ratios. Inset plot shows the first 100 iterations.}
\end{figure}

\subsection{Dynamic Interactions and Convergence}

Finally, we present the convergence results for Algorithm 1. First, in Figure \ref{fig:convergence-lattice}
we plot the value over time of the sharing ratio (or, sharing ratio)
$\rho_{i}(t)=\bar{R}_{i}(t)/D_{i}$ for three nodes in the lattice
network of Figure \ref{fig:lattice-homogeneous}. We observe that
after the first 100 slots (see the inset) the ratios have converged
very closely to their final values. We have plotted the results for
nodes with different degrees, which however does affect in this example
the convergence speed. Similarly, in Figure \ref{fig:convergence-erdos-renyi}
we present the convergence of 4 nodes with different degrees in the
Erdos-Renyi network of Figure \ref{fig:E-R-homogeneous-01}. Again
we observe a relatively fast convergence on the final sharing ratios
which are all equal to 1. Finally, Figure \ref{fig:convergence-barabasi}
presents the convergence results for 3 nodes in the Scale-free graph
of network \ref{fig:scale-free-1}. In this case, the sharing ratios
converge in different levels, and we show here the results for 3 nodes
attaining 3 different sharing ratio values.

\section{Literature Overview}\label{sec:Related}

The model and questions considered in this work are fundamental and therefore their relation to prior studies in different fields is rather expected. For example, to some extent, similar questions have been investigated in the context of network exchange theory in sociology, the general equilibrium theory in economics, various game theoretic works for networked markets or social networks, and in the context of designing cooperation mechanisms in communication and computing systems. Below we provide a brief overview of the more pertinent results in these areas and explain how our work differs. In particular, one key novel aspect of our paper is that unlike past works where the resources of producers to be distributed were known a priori, here we assume that the resource of the producer is generated dynamically over time and the prosumer allocates the resource chunks as become available. The dynamic policy we propose monitors time averages of offered resources as they evolve to make the allocation decisions.
 
 \subsection{General Equilibrium and Game Theory Models}
 
Our work is directly related to the general equilibrium theory which focuses on the existence of exchange ratio and allocation equilibriums in competitive markets. The first attempt to study such models dates back to 1874 and the efforts of L. Walras \citep{walras} who introduced the tatonnement process that leads to an equilibrium, and the analysis of I. Fisher who considered a simpler model (the Fisher market) in 1891 \citep{fisher} and studied automated methods for computing the equilibriums. A richer production - consumption macroeconomic model was introduced in the seminal paper of K. Arrow and G. Debreu \citep{arrow-debreu} who formally proved the existence of equilibriums (A-D model). Subsequent works refined these results by devising the necessary and sufficient conditions for the existence of equilibriums when the agents have linear utility functions \citep{gale-1976}. More recently, researchers have focused on the computation complexity of Arrow-Debreu or Fisher equilibriums, and have developed approximate or even exact (pseudo)-polynomial algorithms for special cases of these markets \citep{devanur-vazirani-2008,orlin-stoc-2010,mehlhorn-2015}; we refer the reader to \citep{devanur-2016} for an excellent discussion about these results and overview of the recent literature.
 
The sharing economy model in this paper differs from the above works in many ways. First, we do not assume the existence of any monetary instrument; hence commodity prices cannot be introduced nor we need to calculate them in order to find the economy's equilibriums. In other words, this is a pure exchange or bartering scheme. Similar models have been studied for house exchange markets \citep{shapley-scarf} or labor (timeshare) exchanges \citep{krishna06}, where the focus has been again to prove existence of equilibriums. Another distinct aspect of our model is the underlying graph that constraints the possible trades. It is worthwhile to note here that A-D and Fisher markets can capture to some extent the network constraints through the commodity preferences (assuming that each agent possesses a different commodity). However, even under this assumption, previous works did not study the impact of the preference constraints on the equilibriums. Similar network models have been studied in the context of graphical economies \cite{KearnsEconSocial2004,KearnsGraphEcon2004} which extend the classical Walrasian equilibrium by imposing constraints on the subsets of buyers and sellers who can trade. It is shown that this network structure can create variation in the price of a certain commodity across different graph neighborhoods. 
 
Compared to these latter results, our model does not presume any type of money transfers, there is no budget constraints (as in typical exchange economies) and the nodes do not value money. Besides, we fully characterize the equilibriums, relate them to the max-min fair solution, and study how they are affected by the network. We also prove that these sharing equilibriums lie within the core of the respective NTU game. Although this relation is known for market games and the respective coalitional games, to the best of our knowledge this is the first result for NTU coalitional graph-constrained games without money. This property is also related to \textit{strong} Nash equilibriums (see \citep{strong-nash} and references therein), for which however there are no general existence results; besides, we focus on market clearing (or competitive) equilibriums which rely on the assumption of price-taking behavior for the agents and hence are fundamentally different than Nash equilibriums. Finally, \citep{Herings2000} and \citep{JacksonWolinsky1996} studied also core solutions of coalitional graph games where the nodes are allowed to create new or severe existing connections. In our model the graph is exogenously given, e.g., based on the location of nodes. Moreover, we go a further step and prove that these equilibriums are strongly stable in the sense that there are no deviations even if they improve only one user's sharing ratio. However, probably the most important difference of our work compared to the above literature is that we explicitly focus on dynamic policies that can lead to the market equilibrium in a decentralized fashion, with lightweight computations and minimal network state information. Namely, the generated resources are unknown in advance and each node is aware only of the existence and strategy of its one-hop neighbors.
 
\subsection{Dynamic Sharing Algorithms}
 
Indeed, the vast majority of literature in Arrow-Debreu or Fisher market (and in their variants) focuses either on the existence of equilibriums or on centralized algorithms for their computation. Albeit very significant, such solutions cannot be applied in practice, especially in the sharing economy applications we focus in this work. Instead, it is of utmost importance to understand if (and how) the agents in such economies can make allocation decisions that will gradually drive the system to its equilibrium, without the need for a central coordinator or full knowledge of the market's state. In \citep{zhang-proportional} the authors studied exactly this problem and proposed a new distributed algorithm, called the Proportional Response (PR) dynamics, as a protocol for trading bandwidth on a peer-to-peer file sharing network. The PR dynamics involves a sequence of bids by the buyers and exchange ratios by the sellers that respond to each other. The seller exchange ratios are simply set to be the sum of all the bids they receive. The buyers set their bids proportional to the utility they would obtain with the bids and the exchange ratios in the previous round. Subsequently, \citep{zhang-dynamics} studied the application of these dynamics for  trading in a Fisher market, while \citep{devanur-dynamics-EC11} showed that the proportional bidding and allocation algorithm introduced in \citep{zhang-proportional} is essentially a gradient descent algorithm that can solve the Fisher market optimization program in a decentralized fashion.
 
This proportional allocation algorithm is very interesting and substantially different from the Walrasian tatonnement process. In the latter the exchange ratio of each good is gradually adjusted according to the excess of demand in the previous time step, and then the agents adjust their requested amounts based on the updated exchange ratios. The proportional response dynamics on the other hand do not explicitly involve a exchange ratio mechanism as the requests are based on the user's utility. Consequently, it does not presume a central controller to aggregate all the demands and offers, and it does not require to solve an optimization problem at each step. However, all the above approaches refer to a static model, where the amounts of commodities and budgets are constant and known in advance. In this work we consider a more challenging (and realistic) system where the resource availability varies randomly and we prove the convergence of a distributed algorithm that fully allocates the generated commodities to the requester that offers the highest sharing ratio. This decision rule is intuitive and in line with the expected market behavior of agents who myopically seek to maximize their benefits in each step. Moreover, we explain that the propose algorithm converges to the equilibrium even if each agent does not know in advance his average expected rate of resource generation, and simply uses the running average in each step. Finally, previous works focused on Fisher markets, where there are only buyers who compete for a set of limited commodities, i.e., the exchange ratios reflect only the congestion and not the interests of the sellers, a condition substantially different from the sharing economy market of prosumers that we study here.
 
\subsection{Cooperation Mechanisms in Communication Systems }
 
The problem of cooperation in wireless networks is of paramount importance and has been considered in different contexts, such as for ad hoc networks \citep{hubaux-coop} or WiFi sharing models \citep{efstathiou}. This is a problem that gains increasing interest in communication networks \citep{Sofia-UPN}, as there are many related market solutions offered either as a product, e.g., routers that are tailored for cooperation \citep{FON}, or mobile applications that allow sharing of content among devices. Moreover, such mechanisms have been extensively studied in file sharing peer-to-peer (P2P) overlay systems. In this case, each device that participates in the overlay is an economic agent in the sense that it provides some commodity (the files it possesses) and requests some other goods (the files other devices share). While our model is similar to previous works, e.g., see \citep{RJohariToNBilateral2011} and references therein, our analysis provides novel insights for the structure and properties of the resulting equilibriums and focuses, for the first time, in the dynamics of such interactions. Finally, in more grassroots efforts, such cooperative infrastructures offer low-cost Internet access to under-served or low-income communities around the globe \citep{redhook}. Unlike previous works, our model does not presume any kind of logistics infrastructure, e.g., for transaction or reputation systems. Instead, we proved that a simple and intuitive best response algorithm, with no information about the graph and resource endowments, converges to a fair and robust connectivity or content sharing equilibrium.

\subsection{Sharing Economy}
 
Sharing economy is a recent topic and hence there are only few related studies on the underlying exchange mechanisms. A first important research thrust here consists of works which analyze the users motives for participation through surveys. It is found that users join various sharing economy platforms often due to sustainability concerns \citep{Ham15,La15} and their decisions are facilitated when third-party mediators (as in Airbnb) \citep{Kim15,IK14} resolve the various logistic and security issues that arise. In \citep{Zev16} it is shown that sharing economy services have non-negligible impact on incumbent businesses, motivating the study of mixed markets where both sharing economy and mainstream services coexist \citep{Cus15}. In sharing systems, reciprocity mechanisms are crucial \citep{Suh10,La13,YE16} and they are affected by the graph structure, an aspect that was first pointed in the context of network exchange theory \citep{Markv88,Bon93,Wlk00}. However, these latter works focus only on simple bargaining problems and their experimental behavioral analysis is orthogonal to our efforts.
 
\section{Concluding Remarks} \label{sec:conclusions}

We introduced and analyzed a novel model of sharing economy networks where agents embedded in a graph share their resources over time. This basic model captures an increasing number of business cases where end-users exchange directly resources or services. Such solutions are fueled today by the ideas about collaborative consumption, or sharing economy, and have the potential to boost the global economy in many ways, transforming the way humans trade and collaborate. Our findings suggest that there is a simple class of dynamic exchange policies that can lead to an equilibrium point which is fair, in a max-min fashion, and stable with respect to selfish strategies of single or groups of colluding agents. Moreover, the equilibrium allocations have an interesting structure that underline the relation of the sharing equilibriums with the topology of the network graph.

Our model can be substantially extended, covering this way an even larger set of scenarios. First, the presented policy is distributed since each node needs to be informed only about the sharing ratios of its outgoing neighbors. As it stands now however, each node needs to know its average resource generation rate $D_{i}$. Nevertheless, this requirement can also be removed by replacing $D_{i}(t)$ with
\[
\bar{{D}(t)}=\frac{\sum_{\tau=1}^{t}D_{i}\left(\tau\right)}{t}\,.
\]
The arguments for proving this claim are mainly technical, albeit lengthy; hence we refer the interested reader to \cite{tsoucas-MOR}
where a different system with similar dynamics is analyzed. In practice, this means that the proposed algorithm can drive the sharing economy network to the desirable operation point with minimum local interactions among the nodes, and with no information about the actual resource availability (statistics of resource generation).%

Another issue with Algorithm 1 is that if the statistics of node endowments change, then decisions are based on time averages, the adaptation of the system to new parameters will be slow. This issue can be avoided by replacing time averages with their discounted versions, e.g. by replacing $\bar{{D}_{i}(t)}$ with $\hat{D}_{i}(t)=(1-\alpha)D_{i}(t)+\alpha\hat{D}_{i}(t-1),t\geq2,\ \hat{D}_{i}(1)=D_{i}(1),\,\,0<\alpha<1$. It can be shown that by selecting $\alpha$ close to 1, the system performance is close to the optimal, while ensuring satisfactory speed of adaptation to statical changes of parameters. We refer the reader to \citep{tsoucas-MOR} for more details on this. Finally, we have assumed that all the users have the same priority and unsatiated demand; the model can be extended for users with different priorities.

%
%


%
%
%

\section*{Appendix} 


\subsection{Proof of Lemma \ref{lem:inequality-1}}

\textbf{Lemma 1}. Under any policy $\pi\in\Pi$ it holds for
	any $\mathcal{{S}}\subseteq\mathcal{{N}},$ 
	\begin{align}
	\limsup_{t\rightarrow\infty}\sum_{i\in{\mathcal{S}}}\bar{R}_{i}^{\pi}\left(t\right) & \leq f({\mathcal{S}}),\\
	\sum_{i\in{\mathcal{S}}}r_{i}^{\pi} & \leq f\left({\mathcal{S}}\right).
	\end{align}
\vspace{2mm}

\textbf{Proof}. According to the definitions and eq. (\ref{eq:resource-allocated}):
\begin{align*}
\sum_{i\in\mathcal{S}}\bar{R}_{i}^{\pi}\left(t\right)=\frac{\sum_{i\in\mathcal{S}}\sum_{\tau=1}^{t}\sum_{j\in\mathcal{N}_{i}}R_{ji}\left(\tau\right)}{t}=\frac{\sum_{j\in\mathcal{\mathcal{N}_{\mathcal{S}}}}\sum_{\tau=1}^{t}\sum_{i\in\mathcal{N}_{j}}R_{ji}\left(\tau\right)}{t}\leq\frac{\sum_{j\in\mathcal{\mathcal{N}_{\mathcal{S}}}}\sum_{\tau=1}^{t}D_{j}(\tau)}{t}.
\end{align*}
Taking limits and using the fact that $\lim_{t\rightarrow\infty}\left(\sum_{\tau=1}^{t}D_{j}(\tau)\right)/t=D_{i},$
(\ref{eq:basic1-2}) follows. Next, 
\begin{align*}
\sum_{i\in{\mathcal{S}}}r_{i}^{\pi}=\sum_{i\in{\mathcal{S}}}\liminf_{t\rightarrow\infty}\bar{R}_{i}^{\pi}\left(t\right)
\leq \liminf_{t\rightarrow\infty}\sum_{i\in{\mathcal{S}}}\bar{R}_{i}^{\pi}\left(t\right) \leq f\left({\mathcal{S}}\right)\mbox{ from (\ref{eq:basic1-2}).}\,\,\,\,\,\blacksquare
\end{align*}


\subsection{Proof of Lemma \ref{lem:f is submodular.-1}}

\textbf{Lemma 2}. $f(\mathcal{{S}})$ is submodular i.e.,it holds for every $\mathcal{S},\ \mathcal{T}\subseteq\mathcal{N},$
	\begin{equation}
	f({\mathcal{S}}\cap{\mathcal{T}})+f\left({\mathcal{S}}\cup{\mathcal{T}}\right)\leq f({\mathcal{S}})+f({\mathcal{T}}).
	\end{equation}

\vspace{2mm}

\textbf{Proof}. By a theorem of Lovasz \citep{lovasz-theorem}, it suffices to show that for any $k\in{\mathcal{N}},$ and ${\mathcal{T}}\subseteq{\mathcal{S}}\subseteq{\mathcal{N}}-\left\{ k\right\} $
	it holds, 
	\[
	f({\mathcal{S}}\cup\left\{ k\right\} )-f\left({\mathcal{S}}\right)\leq f({\mathcal{T}}\cup\left\{ k\right\} )-f\left({\mathcal{T}}\right).
	\]
	Indeed,
	\begin{align*}
	f({\mathcal{T}}\cup\left\{ k\right\} )-f\left({\mathcal{T}}\right) & =\sum_{i\in{\mathcal{N}}_{{\mathcal{T}}\cup\left\{ k\right\} }}D_{i}-\sum_{i\in{\mathcal{N}}_{{\mathcal{T}}}}D_{i}=\sum_{i\in{\mathcal{N}}_{k}-{\mathcal{N}}_{{\mathcal{T}}}}D_{i}\geq\sum_{i\in{\mathcal{N}}_{k}-{\mathcal{N}}_{{\mathcal{\mathcal{S}}}}}D_{i}\ \mbox{ since \ensuremath{{\mathcal{T}}}}\subseteq{\mathcal{S}}\\
	& =f({\mathcal{S}}\cup\left\{ k\right\} )-f\left({\mathcal{S}}\right).\,\,\,\, \blacksquare
	\end{align*}


\subsection{Proof of Lemma \ref{lem:polybase}}

\textbf{Lemma 3}. It holds: $\mathcal{R}_{0}=\mathcal{A}_{0}.$
\vspace{2mm}

\textbf{Proof}. Observe first that $\mathcal{{R}}_{0}$ is convex since given two policies $\pi_{1}$, $\pi_{2}$ in $\Pi_{0}$, one can design a policy $\pi_{3}\in\Pi_{0}$ with $\boldsymbol{r}^{\pi_{3}}=p\boldsymbol{r}^{\pi_{1}}+\left(1-p\right)\boldsymbol{r}^{\pi_{2}}$, where $p\in[0,1]$, as follows: 
	\[
	D_{ij}^{\pi_{3}}\left(t\right)=pD_{ij}^{\pi_{1}}\left(t\right)+\left(1-p\right)D_{ij}^{\pi_{2}}\left(t\right).
	\]
Since $\mathcal{{R}}_{0}$ is convex and $\mathcal{{R}}_{0}\subseteq\mathcal{{A}}_{0}$ to show that in fact $\mathcal{{R}}_{0}=\mathcal{{A}}_{0}$, it suffices to show that all extreme points of $\mathcal{{A}}_{0}$ belong to	$\mathcal{{R}}_{0}$. Since $\mathcal{{A}}_{0}$ is a base of a polymatroid, its extreme points are defined as follows. Let $\sigma(i)$ be any permutation of node indices. Define also $\mathcal{{S}}_{\sigma}(i)=\{\sigma(1),\ldots,\sigma(i)\}$.	Then an extreme point of $\mathcal{{A}}_{0}$ is the following, 
\begin{align}
r_{\sigma(1)} & =f\left({\mathcal{S}}_{\sigma}(1)\right),\label{eq:alloc1-1}\\
r_{\sigma(i)} & =f({\mathcal{S}}_{\sigma}(i))-f({\mathcal{S}}_{\sigma}(i-1)),\ 2\leq i\leq N,\label{eq:alloc2-1}
\end{align}
and in fact all extreme points of $\mathcal{A}_{0}$ are of the form	(\ref{eq:alloc1-1}), (\ref{eq:alloc2-1}). Consider the policy $\pi^{\sigma}$	that operates as follows: 
	\begin{itemize}
		\item All nodes in $\mathcal{{N}}_{\sigma(1)}$ give always their generated
		resource to $\sigma\left(1\right)$. 
		\item All nodes in $\mathcal{{N}}_{\sigma(k)}-\cup_{l=1}^{k-1}\mathcal{{N}}_{\sigma(l)}$
		give always their generated resource to $\sigma\left(k\right),$ $2\leq k\leq N$. 
	\end{itemize}
It can be easily seen from the definitions that under policy $\pi^{\sigma}$	the long-term resources allocated to the nodes are given by (\ref{eq:alloc1-1}),	(\ref{eq:alloc2-1}). \,\,\,$\blacksquare$

\subsection{Proof of Lemma \ref{lem:levels}}

\textbf{Lemma 5}. Let $\boldsymbol{r}\in\mathcal{A}_{0}$ then: a) If $K=1$ then $v_{1}=1$. b) If $K>1$ , then $v_{1}<1$ and $l_{K}>1$.
\vspace{2mm}

\textbf{Proof}. Since $\boldsymbol{r}\in\mathcal{A}_{0}$, we have
\begin{equation}
\sum_{i\in\mathcal{N}}r_{i}=f\left(\mathcal{N}\right)=\sum_{i\in\mathcal{N}}D_{i}.\label{eq:baseEq}
\end{equation}
	
a) If $K=1$, then $r_{i}=v_{1}D_{i}$ for all $i\in\mathcal{N}$,	hence $v_{1}=1.$
	
b) Let now $K>1$. If $v_{1}\geq1$, then since $v_{k}>v_{1}\ k\geq2$, we have 
\begin{equation}
\sum_{i\in\mathcal{N}}r_{i}=\sum_{k=1}^{K}v_{k}\sum_{i\in\mathcal{\mathcal{L}}_{k}}D_{i}>\sum_{i\in\mathcal{N}}D_{i},
\end{equation}
which contradicts (\ref{eq:baseEq}). Similarly it is shown that $v_{K}>1$. \,\,\,$\blacksquare$


\subsection{Proof of Theorem \ref{thm:Optimality-1} }

\textbf{Theorem 1}. The following hold.
	\begin{itemize}
		\item Policy $\pi^{*}$ is Lexicographically optimal. 
		\item Under $\pi^{*}$the node sharing ratios and long-term received resources converge to the equilibrium sharing ratios and equilibrium received resources. 
		\item Policy $\pi^{*}$ is coalitionally stable.
	\end{itemize}
\vspace{2mm}

\textbf{Proof}. In this subsection we show that the policy $\pi^{*}$ that operates according to Algorithm \ref{alg:1} is lexicographically optimal. The rest of the assertions of the theorem follow from the discussion in Section \ref{subsec:Structure}. The proof is based on stochastic approximation techniques. In particular, we will make use of the following stochastic approximation theorem due to Robbins and Siegmund and follow the approach in \citep{tsoucas-MOR}. 
\begin{theorem}
	\label{thm:Rob-Sig}On a probability space $\left(\Omega,{\mathcal{F}},P\right)$
	equipped with a sequence of $\sigma$-fields ${\mathcal{F}}_{0}\subset\cdots\subset{\mathcal{F}}_{t}\subset{\mathcal{F}}$
	let $z_{t},\ \xi_{t},\ \zeta_{t}$ be non-negative and ${\mathcal{F}}_{t}$-measurable
	random variables such that 
	\[
	E\left[z_{t+1}|{\mathcal{F}}_{t}\right]\leq z_{t}-\zeta_{t}+\xi_{t},\ t=1,\cdots,
	\]
	where 
	\begin{equation}
	\sum_{t=1}^{\infty}\xi_{t}<\infty\ a.s.\label{eq:finitecond}
	\end{equation}
	Then, $\lim_{t\rightarrow\infty}z_{t}$ exists, is finite and $\sum_{t=1}^{\infty}\zeta_{t}<\infty$
	a.s. 
\end{theorem}
We will need the following lemma.
\begin{lemma}
	\label{lem:basic-Lyapunov}Let $J_{i}\left(\boldsymbol{r}\right)=E\left[R_{i}\left(t+1\right)|\bar{\boldsymbol{R}}\left(t\right)=\boldsymbol{r}\right],\ \boldsymbol{r}\geq\boldsymbol{0}.$
	It hodls 
	\begin{equation}
	\sum_{i\in\cup_{k=1}^{n}\mathcal{L}_{k}(\boldsymbol{r})}J_{i}\left(r\right)=f\left(\cup_{k=1}^{n}\mathcal{L}_{k}\left(\boldsymbol{r}\right)\right),\ n=1,\cdots,K\left(\boldsymbol{r}\right).\label{eq:eqbound}
	\end{equation}
	For all $\epsilon>0$ it holds, 
	\begin{equation}
	\inf_{\left\Vert \boldsymbol{r}-\boldsymbol{r}^{*}\right\Vert \geq\epsilon}\ \sum_{i=1}^{N}\frac{1}{D_{i}}\left(r_{i}^{*}-r_{i}\right)\left(J_{i}\left(\boldsymbol{r}\right)-r_{i}\right)>0.\label{eq:Positivity}
	\end{equation}
\end{lemma}
\textbf{Proof.} According to Algorithm \ref{alg:1}, the nodes in $\mathcal{L}_{1}\left(\boldsymbol{r}\right)$ receive all the resources generated by their neighbor nodes at time	$t+1$. Similarly, the nodes in $\cup_{k=1}^{n}\mathcal{L}_{n}\left(\boldsymbol{r}\right)$ receive all the resources generated by their neighbors at time $t+1$, hence, 
\[
\sum_{i\in\cup_{n=1}^{k}\mathcal{L}_{n}\left(\boldsymbol{r}\right)}R_{i}\left(t+1\right)=\sum_{i\in\cup_{n=1}^{k}\mathcal{L}_{n}\left(\boldsymbol{r}\right)}D_{i}\left(t+1\right).
\]
Taking conditional expectations we obtain 
\[
\sum_{i\in\cup_{k=1}^{n}\mathcal{L}_{1}\boldsymbol{r}}J_{i}\left(r\right)=\sum_{i\in\cup_{k=1}^{n}\mathcal{L}_{n}\left(\boldsymbol{r}\right)}D_{i}=f\left(\cup_{k=1}^{n}\mathcal{L}_{k}\left(\boldsymbol{r}\right)\right).
\]

To show (\ref{eq:Positivity}), setting $C=\min_{i=1,...,N}\left\{ 1/D_{i}\right\} $ we have for $\left\Vert \boldsymbol{r}-\boldsymbol{r}^{*}\right\Vert \geq\epsilon$,
\begin{align*}
\sum_{i=1}^{N}\frac{1}{D_{i}}\left(r_{i}^{*}-r_{i}\right)\left(J_{i}\left(\boldsymbol{r}\right)-r_{i}\right) & =\sum_{i=1}^{N}\frac{1}{D_{i}}\left(r_{i}^{*}-r_{i}\right)\left(J_{i}\left(\boldsymbol{r}\right)-r_{i}^{*}\right)+\sum_{i=1}^{N}\frac{1}{D_{i}}\left(r_{i}^{*}-r_{i}\right)^{2}\\
& \geq\sum_{i=1}^{N}\frac{1}{D_{i}}r_{i}^{*}\left(J_{i}\left(\boldsymbol{r}\right)-r_{i}^{*}\right)-\sum_{i=1}^{N}\frac{1}{D_{i}}r_{i}\left(J_{i}\left(\boldsymbol{r}\right)-r_{i}^{*}\right)+\epsilon^{2}C.
\end{align*}
Hence to prove (\ref{eq:Positivity}) it suffices to show that 
\begin{equation}
\sum_{i=1}^{N}\frac{1}{D_{i}}r_{i}\left(J_{i}\left(\boldsymbol{r}\right)-r_{i}^{*}\right)\leq0,\label{eq:first}
\end{equation}
and 
\begin{equation}
\sum_{i=1}^{N}\frac{1}{D_{i}}r_{i}^{*}\left(J_{i}\left(\boldsymbol{r}\right)-r_{i}^{*}\right)\geq0.\label{eq:second}
\end{equation}
Next we describe the structure of the lexicographically optimal vector in $\mathcal{A}$. To show (\ref{eq:first}) write , %
\begin{align}
& \sum_{i=1}^{N}\frac{1}{D_{i}}r_{i}\left(J_{i}\left(\boldsymbol{r}\right)-r_{i}^{*}\right)=\sum_{k=1}^{K\left(\boldsymbol{r}\right)}v_{k}
\left(\boldsymbol{r}\right)\sum_{i\in\mathcal{L}_{k}\left(\boldsymbol{r}\right)}\left(J_{i}\left(\boldsymbol{r}\right)-r_{i}^{*}\right)\\ \nonumber
&=\sum_{k=1}^{K\left(\boldsymbol{r}\right)-1}\left(\sum_{n=k}^{K\left(\boldsymbol{r}\right)-1}\left(v_{n}\left(\boldsymbol{r}\right)-v_{n+1}
\left(\boldsymbol{r}\right)\right)+v_{K\left(\boldsymbol{r}\right)}\left(r\right)\right)\left(\sum_{i\in\mathcal{L}_{k}\left(\boldsymbol{r}\right)}J_{i}
\left(\boldsymbol{r}\right)-\sum_{i\in\mathcal{L}_{k}\left(\boldsymbol{r}\right)}r_{i}^{*}\right)\\ \nonumber
&+v_{K\left(\boldsymbol{r}\right)}\left(\boldsymbol{r}\right)\left(\sum_{i\in\mathcal{L}_{K\left(\boldsymbol{r}\right)}\left(\boldsymbol{r}\right)}J_{i}\left(\boldsymbol{r}\right)
-\sum_{i\in\mathcal{L}_{K\left(\boldsymbol{r}\right)}\left(\boldsymbol{r}\right)}r_{i}^{*}\right)\\ \nonumber
&=\sum_{k=1}^{K\left(\boldsymbol{r}\right)-1}\left(\sum_{n=k}^{K\left(\boldsymbol{r}\right)-1}\left(v_{n}\left(\boldsymbol{r}\right)-
v_{n+1}\left(\boldsymbol{r}\right)\right)\right)\left(\sum_{i\in\mathcal{L}_{k}\left(\boldsymbol{r}\right)}J_{i}\left(\boldsymbol{r}\right)-
\sum_{i\in\mathcal{L}_{k}\left(\boldsymbol{r}\right)}r_{i}^{*}\right)+v_{K\left(\boldsymbol{r}\right)}\left(\boldsymbol{r}\right)\left(\sum_{i\in\mathcal{N}}J_{i}
\left(\boldsymbol{r}\right)-\sum_{i\in\mathcal{N}}r_{i}^{*}\right)\\ \nonumber
& =\sum_{n=1}^{K\left(\boldsymbol{r}\right)-1}\left(v_{n}\left(\boldsymbol{r}\right)-v_{n+1}\left(\boldsymbol{r}\right)\right)\left(\sum_{i\in\cup_{k=1}^{n}\mathcal{L}_{k}\left(\boldsymbol{r}\right)}J_{i}\left(\boldsymbol{r}\right)-\sum_{i\in\cup_{k=1}^{n}\mathcal{L}_{k}\left(\boldsymbol{r}\right)}r_{i}^{*}\right)+v_{K\left(\boldsymbol{r}\right)}\left(\boldsymbol{r}\right)\left(\sum_{i\in\mathcal{N}}J_{i}\left(\boldsymbol{r}\right)-\sum_{i\in\mathcal{N}}r_{i}^{*}\right)\\ \nonumber
& =\sum_{n=1}^{K\left(\boldsymbol{r}\right)-1}\left(v_{n}\left(\boldsymbol{r}\right)-v_{n+1}\left(\boldsymbol{r}\right)\right)\left(f\left(\cup_{k=1}^{n}\mathcal{L}_{k}\left(\boldsymbol{r}\right)\right)-\sum_{i\in\cup_{k=1}^{n}\mathcal{L}_{k}\left(\boldsymbol{r}\right)}r_{i}^{*}\right)+v_{K\left(\boldsymbol{r}\right)}\left(\boldsymbol{r}\right)\left(f\left(\mathcal{N}\right)-\sum_{i\in\mathcal{N}}r_{i}^{*}\right)\ \text{by (}\ref{eq:eqbound})\\ \nonumber
& \leq0,\ \text{since \ensuremath{\boldsymbol{r}^{*}\in{\mathcal{R}}_{0}} and \ensuremath{v_{n}\left(\boldsymbol{r}\right)\leq v_{n+1}\left(\boldsymbol{r}\right),\ n=1,\cdots,K\left(\boldsymbol{r}\right)-1.}}
\end{align}
To show (\ref{eq:second}) we repeat essential the same procedure but we sum over the indices $l_{k}^{\boldsymbol{r}^{*}},\ k=1,\cdots,K^{\boldsymbol{r}^{*}}$.\,\,\,$\blacksquare$

We now proceed to show that $\pi^{*}$ is lexicographically optimal, i.e., 
\[
\lim_{t\rightarrow\infty}\bar{R}_{i}\left(t\right)=\boldsymbol{r}^{*}.
\]
Write 
\begin{align}
\bar{R}_{i}\left(t+1\right) & =\frac{\sum_{\tau=1}^{t+1}R_{i}\left(t\right)}{t+1}=\frac{t}{t+1}\frac{\sum_{\tau=1}^{t}R_{i}\left(t\right)}{t}+\frac{R_{i}\left(t+1\right)}{t+1}\nonumber \\
& =\frac{t}{t+1}\bar{R}_{i}\left(t\right)+\frac{R_{i}\left(t+1\right)}{t+1}=\bar{R}_{i}\left(t\right)+\frac{1}{t+1}\left(R_{i}\left(t+1\right)-\bar{R}_{i}\left(t\right)\right),\label{eq:recursion}
\end{align}
and consider the Lyapunov function, 
\[
V\left(\boldsymbol{x}\right)=\frac{1}{2}\sum_{i=1}^{N}\frac{1}{D_{i}}\left(x_{i}-r_{i}^{*}\right)^{2}.
\]
From (\ref{eq:recursion}) we obtain, 
\begin{align*}
&V\left(\bar{\boldsymbol{R}}\left(t+1\right)\right)  =\frac{1}{2}\sum_{i\in\mathcal{N}}\frac{1}{D_{i}}\left(\bar{R}_{i}\left(t\right)+\frac{1}{t+1}\left(R_{i}\left(t+1\right)-\bar{R}_{i}\left(t\right)\right)-r_{i}^{*}\right)^{2}\\
& =V\left(\bar{\boldsymbol{R}}\left(t\right)\right)-\frac{1}{t+1}\sum_{i\in\mathcal{N}}\frac{1}{D_{i}}\left(R_{i}\left(t+1\right)-\bar{R}_{i}\left(t\right)\right)\left(r_{i}^{*}-\bar{R}_{i}\left(t\right)\right)+\frac{1}{2\left(t+1\right)^{2}}\sum_{i\in\mathcal{N}}\frac{1}{D_{i}}\left(R_{i}\left(t+1\right)-\bar{R}_{i}\left(t\right)\right)^{2}.
\end{align*}
Apply now Theorem \ref{thm:Rob-Sig} with with $z_{t}=V\left(\bar{\boldsymbol{R}}\left(t\right)\right),$
\begin{align*}
\zeta_{t} & =\frac{1}{t+1}E\left[\sum_{i\in\mathcal{N}}\frac{1}{D_{i}}\left(R_{i}\left(t+1\right)-\bar{R}_{i}\left(t\right)\right)\left(r_{i}^{*}-\bar{R}_{i}\left(t\right)\right)\left|\bar{R}_{i}\left(t\right)\right.\right]\\
& =\frac{1}{t+1}\sum_{i\in\mathcal{N}}\frac{1}{D_{i}}\left(J_{i}\left(\bar{\boldsymbol{R}}\left(t\right)\right)-\bar{R}_{i}\left(t\right)\right)\left(r_{i}^{*}-\bar{R}_{i}\left(t\right)\right),
\end{align*}
and 
\[
\xi_{t}=\frac{1}{2\left(t+1\right)^{2}}E\left[\sum_{i\in\mathcal{N}}\frac{1}{D_{i}}\left(R_{i}\left(t+1\right)-\bar{R}_{i}\left(t\right)\right)^{2}\left|\bar{R}_{i}\left(t\right)\right.\right].
\]
Clearly, $z_{t}\geq0$ and $\xi_{t}\geq0$. Also, according to (\ref{eq:Positivity}), $\zeta_{t}\ge0$. Hence the non-negativity of the variables in the theorem holds. Notice that 
\[
\bar{R}_{i}\left(t\right)=\frac{\sum_{\tau=1}^{t}R_{i}\left(\tau\right)}{t}\leq\sum_{j\in\mathcal{N}}\frac{\sum_{\tau=1}^{t}D_{j}\left(\tau\right)}{t}\leq NB,
\]
hence, 
\begin{align*}
\xi_{t} & \leq\frac{1}{\left(t+1\right)^{2}}E\left[\sum_{i\in\mathcal{N}}\frac{1}{D_{i}}\left(R_{i}^{2}\left(t+1\right)+\bar{R}_{i}^{2}\left(t\right)\right)\left|\bar{R}_{i}\left(t\right)\right.\right]\\
& =\frac{1}{\left(t+1\right)^{2}}E\left[\sum_{i\in\mathcal{N}}\frac{1}{D_{i}}R_{i}^{2}\left(t+1\right)\left|\bar{R}_{i}\left(t\right)\right.\right]+\frac{1}{\left(t+1\right)^{2}}\sum_{i\in\mathcal{N}}\frac{1}{D_{i}}\bar{R}_{i}^{2}\left(t\right)\\
& \leq\frac{\hat{B}}{\left(t+1\right)^{2}},
\end{align*}
where 
\[
\hat{B}=N^{2}B^{2}\sum_{i\in\mathcal{N}}\frac{1}{D_{i}}.
\]
It follows that $\lim_{t\rightarrow\infty}\sum_{t=1}^{\infty}\xi_{t}<\infty,$
i.e. (\ref{eq:finitecond}) holds. Hence, according to Theorem \ref{thm:Rob-Sig}
we have that $\lim_{t\rightarrow\infty}V\left(\bar{\boldsymbol{R}}\left(t\right)\right)$
exists almost surely and that 
\begin{equation}
\sum_{t=1}^{\infty}\frac{1}{t+1}\sum_{i\in\mathcal{N}}\frac{1}{D_{i}}\left(J_{i}\left(\bar{\boldsymbol{R}}\left(t\right)\right)-\bar{R}_{i}\left(t\right)\right)\left(r_{i}^{*}-\bar{R}_{i}\left(t\right)\right)<\infty\ a.s.\label{eq:finite_as}
\end{equation}
We will show next that $\lim_{t\rightarrow\infty}V\left(\bar{\boldsymbol{R}}\left(t\right)\right)=0$
which implies that $\lim_{t\rightarrow\infty}\bar{\boldsymbol{R}}\left(t\right)=\boldsymbol{r}^{*}$
i.e., the policy $\pi^{*}$ is lexicographically optimal.

Assume that $\lim_{t\rightarrow\infty}V\left(\bar{\boldsymbol{R}}\left(t\right)\right)=\alpha>0.$
Since $\left\Vert \bar{\boldsymbol{R}}\left(t\right)-r^{*}\right\Vert \geq2CV\left(\bar{\boldsymbol{R}}\left(t\right)\right),$
where $C=\min_{i\in\mathcal{N}}D_{i},$ we conclude that 
\[
\liminf_{t\rightarrow\infty}\left\Vert \bar{\boldsymbol{R}}\left(t\right)-r^{*}\right\Vert \geq2C\alpha>0,
\]
which by (\ref{eq:Positivity}) implies that 
\[
\liminf_{t\rightarrow\infty}\sum_{i\in\mathcal{N}}\frac{1}{D_{i}}\left(J_{i}\left(\bar{\boldsymbol{R}}\left(t\right)\right)-\bar{R}_{i}\left(t\right)\right)\left(r_{i}^{*}-\bar{R}_{i}\left(t\right)\right)>0
\]
hence 
\[
\sum_{t=1}^{\infty}\frac{1}{t+1}\sum_{i\in\mathcal{N}}\frac{1}{D_{i}}\left(J_{i}\left(\bar{\boldsymbol{R}}\left(t\right)\right)-\bar{R}_{i}\left(t\right)\right)\left(r_{i}^{*}-\bar{R}_{i}\left(t\right)\right)=\infty
\]
which contradicts (\ref{eq:finite_as}).\,\,\,$\blacksquare$

\subsection{Proof of Theorem \ref{thm:MainTh0-1}}

\textbf{Theorem 3}. A vector $\boldsymbol{r}\in\mathcal{\mathcal{A}}$, is lexicographically optimal if and only if the following hold. If $K=1,$ then $v_{1}=1$. If $K\geq2$ then
\begin{enumerate}
\item $\mathcal{L}_{k}$ is an independent set in graph $G_{\mathcal{Q}_{k}}$, for $k=1,....,\lfloor{\frac{K}{2}\rfloor}$. 
\item $\mathcal{L}_{K-k+1}=\mathcal{N}_{\mathcal{Q}_{k}}\left(\mathcal{L}_{k}\right)$, for $k=1,....,\lfloor{\frac{K}{2}\rfloor}$. 
\item $v_{k}v_{K-k+1}=1$, for $k=1,....,\lfloor{K/2\rfloor}$. 
\item $\sum_{i\in\mathcal{L}_{k}}r_{i}=\sum_{i\in\mathcal{L}_{K-k+1}}D_{i}$, for $k=1,....,\lfloor{\frac{K}{2}\rfloor}$. 
\item If $K$ is odd, then $v_{\left\lceil K/2\right\rceil }=1.$
\end{enumerate}
\vspace{2mm}

\textbf{Proof}. Let $K=1.$ Let $\boldsymbol{r}$ be lexicographically optimal, hence $\boldsymbol{r}\in\mathcal{A}_{0}$. Then according to Lemma \ref{lem:levels}, $v_{1}=r_{i}/D_{i}=1,\ i\in\mathcal{N}$. If on the other hand $v_{1}=1$ then
	\[
	\sum_{i\in\mathcal{N}}r_{i}=\sum_{i\in\mathcal{N}}D_{i}=f\left(\mathcal{N}\right),
	\]
and according to Theorem \ref{thm:lextoptvector} $\boldsymbol{r}$ is lexicographically optimal.
	
Next we consider the case $K\geq2.$\\

a) Assume that $\boldsymbol{r}$ is lexicographically optimal and hence
\begin{equation}
\sum_{i\in\mathcal{L}_{1}}r_{i}=f\left(\mathcal{L}_{1}\right)=\sum_{i\in\mathcal{N}_{\mathcal{L}_{1}}}D_{i}.\label{eq:L1Eq}
\end{equation}
We first show that $\mathcal{L}_{1}$ is an independent set. Assume that there are two nodes in $\mathcal{L}_{1}$ that are connected and consider the maximal connected set $\mathcal{L}$ in $\mathcal{L}_{1}$ that contains these two nodes. Fix any allocation set that generates $\boldsymbol{r}$. Since $\mathcal{L}$ is maximal, no node in $\mathcal{L}$ is connected to a node in $\mathcal{L}_{1}$; then, (\ref{eq:L1Eq}) implies that (under the allocation that generates $\boldsymbol{r}$) all nodes in $\mathcal{L}$ give their endowment to nodes in $\mathcal{L}$ and hence, 
\begin{equation}
\sum_{i\in\mathcal{L}}r_{i}\geq\sum_{i\in\mathcal{L}}D_{i},\,\,\, \Rightarrow\,\,\,\,v_{1}\sum_{i\in\mathcal{L}}D_{i}\geq\sum_{i\in\mathcal{L}}D_{i} \nonumber
\end{equation}
or $v_{1}\geq1,$ which contradicts Lemma \ref{lem:levels}.
	
Consider now the smallest level $m$ such that nodes in $\mathcal{L}_{1}$ are connected to some nodes in $\mathcal{L}_{m}$. Let $\mathcal{S}_{1}$ be the set of nodes in $\mathcal{L}_{1}$ which are connected to some nodes in $\mathcal{L}_{m}$. Let $\mathcal{S}_{m}$  be the set of nodes in $\mathcal{L}_{m}$ that are connected to some nodes in $\mathcal{S}_{1}$. Then, since according to Theorem \ref{thm:lextoptvector} the nodes in $\mathcal{S}_{1}$ receive all the endowments of nodes in $\mathcal{S}_{m}$, we have
\begin{equation}
\sum_{i\in\mathcal{S}_{1}}r_{i}\geq\sum_{m\in\mathcal{S}_{m}}D_{m},\,\,\,\mbox{or}\,\,\,\,v_{1}\sum_{i\in\mathcal{S}_{1}}D_{i}\geq\sum_{m\in\mathcal{S}_{m}}D_{m}\,.\label{eq:ineq1-m}
\end{equation}
Also, since nodes in $\mathcal{S}_{1}$ are not connected, the nodes in $\mathcal{S}_{m}$ receive all the endowments of nodes in $\mathcal{S}_{1}$, hence 
\begin{equation}
\sum_{i\in\mathcal{S}_{m}}r_{i}\geq\sum_{i\in\mathcal{S}_{1}}D_{i},\,\,\,\,\mbox{or}\,\,\,\,v_{m}\sum_{i\in\mathcal{S}_{m}}D_{i}\geq\sum_{i\in\mathcal{S}_{1}}D_{i}\,.\label{eq:ineqm-1}
\end{equation}
From (\ref{eq:ineq1-m}) and (\ref{eq:ineqm-1}) and the fact that $\sum_{i\in\mathcal{S}_{m}}D_{i}>0$, $\sum_{i\in\mathcal{S}_{1}}D_{i}>0$, we conclude that 
\begin{equation}
v_{1}v_{K}\geq v_{1}v_{m}\geq1.\label{eq:FirstIneq-1}
\end{equation}
	
Next consider the set $\mathcal{L}_{K}$. Notice first that if node $i$ gives some of its endowment to nodes in $\mathcal{L}_{K}$, then $\mathcal{N}_{i}\in\mathcal{L}_{K}.$ This is so, since if node $i$ has a neighbor in a set $\mathcal{L}_{n},\ n<K$ then according to Theorem \ref{thm:lextoptvector} node $i$ would give all its endowment to nodes with lower levels. It follows that no node in $\mathcal{L}_{K}$ gives endowment to $\mathcal{L}_{K}$. To see this, let $\mathcal{S\neq\emptyset}$ be the set of nodes in $\mathcal{L}_{K}$ that give their endowment to $\mathcal{L}_{K}.$ Then the nodes in $\mathcal{S}$ receive endowments only from nodes in $\mathcal{S}.$ This is so, because if node $i$ gives endowment to $j\in\mathcal{S}$ then $j\in\mathcal{N}_{i}$, hence $j\in\mathcal{L}_{K}$ and then by definition of the set $\mathcal{S},$ $i\in\mathcal{S}.$ Hence, it holds 
\begin{equation}
\sum_{i\in\mathcal{S}}r_{i}\leq\sum_{i\in\mathcal{S}}D_{i},\,\,\,\,\mbox{or}\,\,\,\,\,v_{k}\sum_{i\in\mathcal{S}}D_{i}\leq\sum_{i\in\mathcal{S}}D_{i}\,.
\end{equation}
i.e., $v_{i}\leq1$ which contradicts Lemma \ref{lem:levels}. 
	
Let now $n<K$ be the largest level such that $\mathcal{L}_{n}$ contains a node that gives endowment to $\mathcal{L}_{K}$. Let $\mathcal{S}_{n}$ be the nodes in $\mathcal{L}_{n}$ that give endowment to some node in $\mathcal{L}_{K}.$ Note that the nodes in $\mathcal{S}_{n}$ are connected only to nodes in $\mathcal{S}_{K}$ since otherwise they (the nodes in $\mathcal{S}_{n}$) would give their endowment to nodes at lower levels. Let $\mathcal{S}_{K}$ be the nodes in $\mathcal{N}_{\mathcal{S}_{n}}\cap\mathcal{L}_{K}$ that give endowment to nodes in $\mathcal{S}_{n}.$ Then, since the nodes in $\mathcal{S}_{n}$ are connected only to nodes in $\mathcal{S}_{K}$, it holds 
\begin{equation}
\sum_{i\in\mathcal{S}_{n}}r_{i}\leq\sum_{i\in\mathcal{S}_{K}}D_{i},\,\,\,\,\mbox{or}\,\,\,\,v_{n}\sum_{i\in\mathcal{S}_{n}}D_{i}\leq\sum_{i\in\mathcal{S}_{K}}D_{i}.\label{eq:todown1}
\end{equation}
	
Also, note that the nodes in $\mathcal{S}_{K}$ are not connected to nodes at lower levels than $n$ since otherwise they would give their endowment to these nodes. Since we already showed that these nodes do not get any endowment from nodes in $\mathcal{L}_{K}$, it follows that 
\begin{equation}
\sum_{i\in\mathcal{S}_{K}}r_{i}\leq\sum_{i\in\mathcal{S}_{n}}D_{i},\,\,\,\,\mbox{or}\,\,\,\,\,v_{K}\sum_{i\in\mathcal{S}_{K}}D_{i}\leq\sum_{i\in\mathcal{S}_{n}}D_{i}.\label{eq:todown1-1}
\end{equation}
From (\ref{eq:todown1}), (\ref{eq:todown1-1}) we conclude
\begin{equation}
v_{K}v_{1}\leq v_{K}v_{n}\leq1.\label{eq:secondineq}
\end{equation}
Inequalities (\ref{eq:FirstIneq-1}), (\ref{eq:secondineq}) imply that $v_{K}v_{1}=1,\ m=K$, and $n=1$ and these in turn imply statements of the theorem for $k=1.$ 
	
Next consider the graph $G_{\mathcal{Q}_{2}\left(\boldsymbol{r}\right)}=\left(\mathcal{Q}_{2}\left(\boldsymbol{r}\right),\mathcal{E}_{\mathcal{Q}_{2}}\left(\boldsymbol{r}\right)\right)$ and the vector $\boldsymbol{r}_{2}$ that has components those of vector $\boldsymbol{r}$ that are in $\mathcal{N}-\left(\mathcal{L}_{1}\cup\mathcal{L}_{K}\right)$. It can be easily seen that $\boldsymbol{r}_{2}$ is lexicographically optimal in $G_{\mathcal{Q}_{2}\left(\boldsymbol{r}\right)}$ and therefore we can repeat the process to complete the proof using induction.
	
Assume now that conditions \ref{enu:MainTh0Item1-1}-\ref{enu:main-5} of the theorem hold. Then it can be seen by induction on $k$ that conditions (\ref{eq:lex1}), (\ref{eq:lex2}) of Theorem 2 hold, and hence the vector $\boldsymbol{r}$ is lexicographically optimal. To see this, we describe the case $k=1.$ Since by condition \ref{enu:MainTh0Item1-1} $\mathcal{L}_{1}$ is an independent set, we conclude from conditions \ref{enu:MainTh0Item2-1} and \ref{enu:MainTh0Item3-1} that the nodes in $\mathcal{L}_{1}$ receive all endowments of their neighbors, hence (\ref{eq:lex1}) of Theorem \ref{thm:lextoptvector} is satisfied. Also, 
	\begin{align*}
	\sum_{i\in\mathcal{L}_{K}}r_{i} =v_{K}\sum_{i\in\mathcal{L}_{K}}D_{i}=\frac{1}{v_{1}}\sum_{i\in\mathcal{L}_{1}}r_{i}=\sum_{i\in\mathcal{L}_{1}}D_{i},
	\end{align*}
where the second equality holds due to conditions \ref{enu:MainTh0Item2.1-1} and \ref{enu:MainTh0Item3-1}. The last equality implies that the nodes in $\mathcal{L}_{K}$ receive only the resources of the nodes in $\mathcal{L}_{1}$ who are not connected to nodes of lower level than $K$, hence (\ref{eq:lex2}) is satisfied for $k=K.$ \,\,\,$\blacksquare$

\subsection{Proof of Theorem \ref{thm:competitive}}

\textbf{Theorem 4}. Let $\boldsymbol{r}^{*}$ be a lexicographically optimal vector. The ratios $\left\{ \rho_{i}^{*}\right\} _{i\in\mathcal{N}}=\left\{ r_{i}^{*}/D_{i}\right\} _{i\in\mathcal{N}}$ are equilibrium sharing ratios for the competitive framework. 
\vspace{2mm}

\textbf{Proof}. Notice that $\rho_{i}^{*}=v_{I_{i}}^{*}.$ Since $\boldsymbol{r}^{*}\in\mathcal{A}_{0},$ there is an allocation set $\left\{ d_{ij}^{*}\geq0,\ i\in\mathcal{N},\ j\in\mathcal{N}_{i}\right\}$, such that 
\begin{equation}
\sum_{j\in\mathcal{N}_{i}}d_{ij}^{*}=D_{i},\ i\in\mathcal{N},\label{eq:allocgive-1}
\end{equation}
\begin{equation}
\sum_{j\in\mathcal{N}_{i}}d_{ji}^{*}=r_{i},\ i\in\mathcal{N}.\label{eq:alloctake-1}
\end{equation}
Consider the following policy $\pi^{*}$ that operates as follows. At any time $t$, node $i$ allocates to node $j\in\mathcal{N}_{i}$ resource $D_{ij}\left(t\right)=\frac{d_{ij}^{*}}{D_{i}}$. This implies that under policy $\pi^{*}$ (\ref{eq:exchange0}) is satisfied, i.e., $\lim_{t\rightarrow\infty}\bar{D}(t)=D_{i}$. Moreover, it is easily shown that
\[
\lim_{t\rightarrow\infty}\bar{R}_{i}^{\pi*}\left(t\right)=r_{i}^{*}=D_{i}\rho_{i}^{*}.
\]
Finally, from the structure of $\boldsymbol{r}^{*}$ in Theorem \ref{thm:MainTh0-1} it can be seen by induction on $k$ that every node allocates the resources it generates to neighbors with the smallest sharing ratios.\,\,\,$\blacksquare$

\subsection{Proof of Theorem \ref{thm:stable}}

\textbf{Theorem 5}. A policy $\pi^{*}$ that achieves the lexicographically optimal vector $\boldsymbol{r}^{*}$is strongly stable.  
\vspace{2mm}

\textbf{Proof}. Assume that there is a set $\mathcal{S}\subset\mathcal{N}$ such that the nodes in this set exchange resources only between themselves and achieve rates $\hat{r_{i}}\geq r_{i}^{*},\ i\in\mathcal{S}$ with $\hat{r}_{j}>r_{j}^{*}$ for at least one $j\in\mathcal{S}$. Let $\hat{\pi}$ be the policy that achieves rate vector $\hat{\boldsymbol{r}}$ and let $\hat{d_{ij}},\ i,j\in\mathcal{S}$ be an allocation vector that generates $\hat{\boldsymbol{r}}$, hence,
\[
\sum_{j\in\mathcal{S}\cap\mathcal{N}_{i}}\hat{d}_{ji}=\hat{r}_{i},i\in\mathcal{S}.
\]
We then claim that 
\begin{equation}
r_{i}^{*}<\hat{r}_{i}\Rightarrow\sqrt{\rho_{i}^{*}}D_{i}<\sum_{j\in\mathcal{S}\cap\mathcal{N}_{i}}\sqrt{\rho_{j}^{*}}\hat{d_{ji}},\label{eq:stab1}
\end{equation}
\begin{equation}
r_{i}^{*}\leq\hat{r}_{i}\Rightarrow\sqrt{\rho_{i}^{*}}D_{i}\leq\sum_{j\in\mathcal{S}\cap\mathcal{N}_{i}}\sqrt{\rho_{j}^{*}}\hat{d_{ji}}.\label{eq:stab2}
\end{equation}

To see (\ref{eq:stab1}), assume that $\sqrt{\rho_{i}^{*}}D_{i}\geq\sum_{j\in\mathcal{S}\cap\mathcal{N}_{i}}\sqrt{\rho_{j}^{*}}\hat{d_{ji}}$. Since Theorems \ref{thm:MainTh0-1} and \ref{thm:competitive} imply that 
	\[
	\rho_{j}^{*}\geq\frac{1}{\rho_{i}^{*}},\ j\in\mathcal{N}_{i},
	\]
we have, 
\begin{align}
\sqrt{\rho_{i}^{*}}D_{i} \geq\sum_{j\in\mathcal{S}\cap\mathcal{N}_{i}}\sqrt{\rho_{j}^{*}}\hat{d_{ji}}\geq\frac{1}{\sqrt{\rho_{i}^{*}}}\sum_{j\in\mathcal{S}\cap\mathcal{N}_{i}}\hat{d_{ji}}=\frac{1}{\sqrt{\rho_{i}^{*}}}\hat{r_{i}}.
\end{align}
Therefore, $r_{i}^{*}=\rho_{i}^{*}D_{i}\geq\hat{r_{i}},$ a contradiction. In a similar fashion (\ref{eq:stab2}) can be shown. Summing now (\ref{eq:stab1}), (\ref{eq:stab2}) over $i\in\mathcal{S}$ we have 
\begin{align}
\sum_{i\in\mathcal{S}}\sqrt{\rho_{i}^{*}}D_{i} <\sum_{i\in\mathcal{S}}\ \sum_{j\in\mathcal{S}\cap\mathcal{N}_{i}}\sqrt{\rho_{j}^{*}}\hat{d_{ji}}=\sum_{j\in\mathcal{S}}\ \sqrt{\rho_{j}^{*}}\sum_{i\in\mathcal{S}\cap\mathcal{N}_{j}}\hat{d_{ji}}\leq\sum_{j\in\mathcal{S}}\ \sqrt{\rho_{j}^{*}}D_{j},
\end{align}
a contradiction.\,\,\,$\blacksquare$

\theendnotes



\bibliographystyle{informs2014} 



\end{document}